\documentclass[10pt,twocolumn]{article}
\usepackage[top=1.85cm, bottom=1.85cm, left=1.7cm, right=1.7cm]{geometry}
\usepackage{times}
\usepackage[hyphens]{url}  %
\usepackage[keeplastbox]{flushend}  %
\usepackage{graphicx}  %
\usepackage{xcolor}
\usepackage{flushend}
\usepackage[sort&compress,numbers]{natbib}

\usepackage{booktabs}
\usepackage{graphicx}
\usepackage{paralist}
\usepackage{enumitem}
\usepackage{array}

\usepackage[small,bf]{caption}
\usepackage{caption}
\usepackage{subfigure}

\usepackage[hang,flushmargin]{footmisc}

\usepackage[bookmarks=true,%
            bookmarksnumbered=true,%
            colorlinks=true,%
            linkcolor=linkcol,%
            citecolor=citecol,%
            urlcolor=urlcol,%
            hypertexnames=true,%
            pdfpagelabels,
      ]%
      {hyperref}

\usepackage[OT2,OT1]{fontenc}
\newcommand\cyr
{
\renewcommand\rmdefault{wncyr}
\renewcommand\sfdefault{wncyss}
\renewcommand\encodingdefault{OT2}
\normalfont
\selectfont
}
\DeclareTextFontCommand{\textcyr}{\cyr}

\usepackage[compact]{titlesec}
\titlespacing*{\section}{0pt}{*2.5}{2.5pt} 
\titlespacing{\subsection}{0pt}{*2}{2pt}

\definecolor{linkcol}{rgb}{0,0,0.5}
\definecolor{citecol}{rgb}{0,0.5,0.3}
\definecolor{urlcol}{rgb}{0.3,0,0}

\renewenvironment{thebibliography}[1]{
  \begin{oldthebibliography}{#1}
    \setlength{\itemsep}{0.1em}
    \setlength{\parskip}{0.1em}
}
{
  \end{oldthebibliography}
}

\renewcommand{\footnoterule}{%
  \kern -3pt
  \hrule width 1in
  \kern 2pt
}

\usepackage{xspace}
\newcommand{\descr}[1]{\smallskip\noindent\textbf{#1}}

\makeatletter
\def\url@leostyle{%
  \@ifundefined{selectfont}{\def\UrlFont{}}%
  {\def\UrlFont{}}%
}
\makeatother
\begin{document}
\title{\bf ``How over is it?'' Understanding the Incel Community on YouTube\thanks{To appear at the 24th ACM Conference on Computer-Supported Cooperative Work and Social Computing (CSCW 2021). Please cite the CSCW version.}}

\author{Kostantinos Papadamou$^\star$, Savvas Zannettou$^\mp$, Jeremy Blackburn$^\dagger$\\
Emiliano De Cristofaro$^\ddagger$, Gianluca Stringhini$^\diamond$, Michael Sirivianos$^\star$\\[0.5ex]
\normalsize $^\star$Cyprus University of Technology, $^\mp$Max Planck Institute, $^\dagger$Binghamton University \\[-0.5ex]
\normalsize $^\ddagger$University College London, $^\diamond$Boston University\\
\normalsize ck.papadamou@edu.cut.ac.cy, szannett@mpi\-inf.mpg.de, jblackbu@binghamton.edu \\[-0.5ex]
\normalsize e.decristofaro@ucl.ac.uk, gian@bu.edu, michael.sirivianos@cut.ac.cy
}
\date{}

\maketitle

\begin{abstract}
YouTube is by far the largest host of user-generated video content worldwide. 
Alas, the platform has also come under fire for hosting inappropriate, toxic, and hateful content.
One community that has often been linked to sharing and publishing hateful and misogynistic content are the Involuntary Celibates (Incels), a loosely defined movement ostensibly focusing on men's issues.
In this paper, we set out to analyze the Incel community on YouTube by focusing on this community's evolution over the last decade and understanding whether YouTube's recommendation algorithm steers users towards Incel-related videos.
We collect videos shared on Incel communities within Reddit and perform a data-driven characterization of the content posted on YouTube. 

Among other things, we find that the Incel community on YouTube is getting traction and that, during the last decade, the number of Incel-related videos and comments rose substantially. 
We also find that users have a $6.3\%$ chance of being suggested an Incel-related video by YouTube's recommendation algorithm within five hops when starting from a non Incel-related video.
Overall, our findings paint an alarming picture of online radicalization: not only Incel activity is increasing over time, but platforms may also play an {\em active} role in steering users towards such extreme content.
\end{abstract}

\section{Introduction}
\label{sec:introduction}
While YouTube has revolutionized the way people discover and consume video content online, it has also enabled the spread of inappropriate and hateful content. 
The platform, and in particular its recommendation algorithm, has been repeatedly accused of promoting offensive and dangerous content, and even of helping radicalize users~\cite{youtuberadical_2019,mozila2019RabbitHoleStories,ribeiro2019auditing,tufekci2018youtube}.

One fringe community active on YouTube are the so-called Involuntary Celibates, or \emph{Incels}~\cite{youtubeBansExtremism2019}.
While not particularly structured, Incel ideology revolves around the idea of the ``blackpill'' -- a bitter and painful truth about society -- which roughly postulates that life trajectories are determined by how attractive one is and that things that are largely out of personal control, like facial structure, are more ``valuable'' than those under our control, like the fitness level.
Incels are one of the most extreme communities of the Manosphere~\cite{HowRampageKiller2018}, a larger collection of movements discussing men's issues~\cite{ging2019alphas} (see Section~\ref{sec:background_related_work}).
When taken to the extreme, these beliefs can lead to a dystopian outlook on society, where the only solution is a radical, potentially violent shift towards traditionalism, especially in terms of women's role in society~\cite{incelsLooksmaxing2018}.

Overall, Incel ideology is often associated with misogyny and anti-feminist viewpoints, and it has also been linked to multiple mass murders and violent offenses~\cite{malesupr2019splc,fifthestate2019cbc}.
In May 2014, Elliot Rodger killed six people and himself in Isla Vista, CA.
This incident was a harbinger of things to come.
Rodger uploaded a video on YouTube with his ``manifesto,'' as he planned to commit mass murder due to his belief in what is now generally understood to be Incel ideology~\cite{elliotrodgers_2014}.
He served as an apparent ``mentor'' to another mass murderer who shot nine people at Umpqua Community College in Oregon the following year~\cite{harpermercer_2015}.
In 2018, another mass murderer drove his van into a crowd in Toronto, killing nine people, and after his interrogation, the police claimed he had been radicalized online by Incel ideology~\cite{cecco_2019}. 
More recently, 22-year-old Jake Davison shot and killed five people, including a 3-year-old girl, in Plymouth, England~\cite{plymouth}.
Thus, while the concepts underpinning Incels' principles may seem absurd, they also have grievous real-world consequences~\cite{hoffman2020assessing,insideincels2018washington,incel2018vox}.

\descr{Motivation.} 
Online platforms like Reddit became aware of the problem and banned several Incel-related communities on the platform~\cite{incelssubredditbanned_2017}.
However,  prior work suggests that banning subreddits and their users for hate speech does not solve the problem, but instead makes these users someone else's problem~\cite{chandrasekharan2017you}, as banned communities migrate to other platforms~\cite{newell2016user}.
Indeed, the Incel community comprising several banned subreddits ended up migrating to various other online communities such as new subreddits, stand-alone forums, and YouTube channels~\cite{ribeiro2020pick,manoel_bans}.

The research community has mostly studied the Incel community and the broader Manosphere on Reddit, 4chan, and online discussion forums like Incels.me or Incels.co~\cite{farrell2019exploring,nagle2015investigation,jaki2019online,ribeiro2020pick,maxwell2020short,manoel_bans}.
However, the fact that YouTube has been repeatedly accused of user radicalization and promoting offensive and inappropriate content~\cite{youtuberadical_2019,mozila2019RabbitHoleStories,ribeiro2019auditing,rabbithole2019donovan,kaiser2018unite} prompts the need to study the extent to which Incels are exploiting the YouTube platform to spread their views.

\descr{Research Questions.} 
With this motivation in mind, this paper explores the footprint of the Incel community on YouTube. 
More precisely, we identify two main research questions:
\begin{enumerate}
\item \textbf{RQ1:} How has the Incel community evolved on YouTube over the last decade?
\item \textbf{RQ2:} Does YouTube's recommendation algorithm contribute to steering users towards Incel communities? 
\end{enumerate}

\descr{Methods.}
We collect a set of 6.5K YouTube videos shared on Incel-related subreddits (e.g., /r/incels, /r/braincels, etc.), as well as a set of 5.7K random videos as a baseline.
We then build a lexicon of 200 Incel-related terms via manual annotation, using expressions found on the Incel Wiki.
We use the lexicon to label videos as {\em ``Incel-related,''} based on the appearance of terms in the transcript, which describes the video’s content, and comments on the videos. 
Next, we use several tools, including temporal and graph analysis, to investigate the evolution of the Incel community on YouTube and whether YouTube's recommendation algorithm contributes to steering users towards Incel content.
To build our graphs, we use the YouTube Data API, which lets us analyze YouTube's recommendation algorithm’s output based on video item-to-item similarities, as well as general user engagement and satisfaction metrics~\cite{zhao2019recommending}.

\descr{Main Findings.} 
Overall, our study yields the following main findings:
\begin{itemize}
  \item We find an increase in Incel-related activity on YouTube over the past few years and in particular concerning Incel-related videos, as well as comments that include pertinent terms. This indicates that Incels are increasingly exploiting the YouTube platform to broadcast and discuss their views.

  \item Random walks on the YouTube's recommendation graph using the Data API and without personalization reveal that with a $6.3\%$ probability a user will encounter an Incel-related video within five hops if they start from a random non-Incel-related video posted on Reddit. Simultaneously, Incel-related videos are more likely to be recommended within the first two to four hops than in the subsequent hops. 

  \item We also find a $9.4\%$ chance that a user will encounter an Incel-related video within three hops if they have visited Incel-related videos in the previous two hops. This means that a user who purposefully and consecutively watches two or more Incel-related videos is likely to continue being recommended such content and with higher frequency.
\end{itemize}
Overall, our findings indicate that Incels are increasingly exploiting YouTube to spread their ideology and express their misogynistic views. 
They also indicate that the threat of recommendation algorithms nudging users towards extreme content is real and that platforms and researchers need to address and mitigate these issues. 

\descr{Paper Organization.}
We organize the rest of the paper as follows. The next section presents an overview of Incel ideology and the Manosphere and a review of the related work. 
Section~\ref{sec:data_collection} provides information about our data collection and video annotation methodology, while Section~\ref{sec:temporal_analysis} analyzes the evolution of the Incel community on YouTube.
Section~\ref{sec:recommendation_analysis} presents our analysis of how YouTube's recommendation algorithm behaves with respect to Incel-related videos.
Finally, we discuss our findings and possible design implications for social media platforms, and conclude the paper in Section~\ref{sec:discussion}.

\section{Background \& Related Work}
\label{sec:background_related_work}

Incels are a part of the broader ``Manosphere,'' a loose collection of groups revolving around a common shared interest in ``men's rights'' in society~\cite{ging2019alphas}. 
While we focus on Incels, understanding the overall Manosphere movement provides relevant context.
In this section, we provide background information about Incels and the Manosphere.
We also review related work focusing on understanding Incels on the Web, YouTube's recommendation algorithm and user radicalization, as well as harmful activity on YouTube.

\subsection{Incels and the Manosphere}\label{sec:mano}
\descr{The Manosphere.} 
The emergence of the so-called Web 2.0 and popular social media platforms have been crucial in enabling the Manosphere~\cite{marwick2017media}.
Although the Manosphere had roots in anti-feminism~\cite{messner1998limits,farrell1996myth}, it is ultimately a \emph{reactionary} community, with its ideology evolving and spreading mainly on the Web~\cite{ging2019alphas}. 
Blais et al.~\cite{blais2012masculinism} analyze the beliefs concerning the Manosphere from a sociological perspective and refer to it as masculinism.
They conclude that masculinism is: ``a trend within the anti-feminist counter-movement mobilized not only against the feminist movement but also for the defense of a non-egalitarian social and political system, that is, patriarchy.'' 
Subgroups within the Manosphere actually differ significantly.  For instance, Men Going Their Own Way (MGTOWs) are hyper-focused on a particular set of men's rights, often in the context of a bad relationship with a woman. These subgroups should not be seen as distinct units. Instead, they are interconnected nodes in a network of misogynistic discourses and beliefs~\cite{bratich2019pickup}. According to Marwick and Lewis~\cite{marwick2017media}, what binds the manosphere subgroups is ``the idea that men and boys are victimized; that feminists, in particular, are the perpetrators of such attacks.''

Overall, research studying the Manosphere has been mostly theoretical and qualitative in nature~\cite{hunteFemaleNatureCucks,ging2019alphas,linAntifeminismOnlineMGTOW2017,gotell2016SexualViolenceManospherea}. 
These qualitative studies are important because they guide our study in terms of framework and conceptualization while motivating large-scale data-driven work like ours.

\descr{Incels.} 
Incels are arguably the most extreme subgroup of the Manosphere~\cite{HowRampageKiller2018}.
Incels appear disarmingly honest about what is causing their grievances compared to other radical ideologies. They openly put their sexual deprivation, which is supposedly caused by their unattractive appearance, at the forefront, thus rendering their radical movement potentially more persuasive and insidious~\cite{maxwell2020short}.
Incel ideology differs from the other Manosphere subgroups in the significance of the ``involuntary'' aspect of their celibacy. They believe that society is rigged against them in terms of sexual activity, and there is no solution at a personal level for the systemic dating problems of men~\cite{incelwikiBlackpillIncelWiki,rationalwikiIncelRationalWiki,menzie2020stacys}.
Further, Incel ideology differs from, for example, MGTOW, in the idea of \emph{voluntary} vs. \emph{involuntary} celibacy. MGTOWs are \emph{choosing} to not partake in sexual activities, while Incels believe that society adversarially deprives them of sexual activity. This difference is crucial, as it gives rise to some of their more violent tendencies~\cite{ging2019alphas}.

Incels believe to be doomed from birth to suffer in a modern society where women are not only able but encouraged to focus on superficial aspects of potential mates, e.g., facial structure or racial attributes. Some of the earliest studies of ``involuntary celibacy'' note that celibates tend to be more introverted and that, unlike women, celibate men in their 30s tend to be poorer or even unemployed~\cite{kiernan1988remains}.
In this distorted view of reality, men with these desirable attributes (colloquially nicknamed {\em Chads} by Incels) are placed at the top of society's hierarchy.
While a perusal of influential people in the world would perhaps lend credence to the idea that ``handsome'' white men are indeed at the top, the Incel ideology takes it to the extreme.

Incels rarely hesitate to call for violence~\cite{baele2019saint}. 
For example, when they seek advice from other Incels about their physical appearance using the phrase ``How over is it?,'' they may be encouraged to ``rope'' (to hang oneself)~\cite{vice2018incellangdecode}.
Occasionally they call for outright gendercide.
Zimmerman et al.~\cite{zimmerman2018recognizing} associate Incel ideology to white-supremacy, highlighting how it should be taken as seriously as other forms of violent extremism.

\subsection{Related Work}
\descr{Incels and the Web.} 
Massanari~\cite{massanari2017gamergate} performs a qualitative study of how Reddit's algorithms, policies, and general community structure enables, and even supports, toxic culture. She focuses on the \#GamerGate and Fappening incidents, both of which had primarily female victims, and argues that specific design decisions make it even worse for victims. For instance, the default ordering of posts on Reddit favors mobs of users promoting content over a smaller set of victims attempting to have it removed.
She notes that these issues are exacerbated in the context of online misogyny because many of the perpetrators are extraordinarily techno-literate and thus able to exploit more advanced features of social media platforms.

Baele et al.~\cite{baele2019saint} study content shared by members of the Incel community, 
focusing on how support and motivation for violence result from their worldview.
Farell et al.~\cite{farrell2019exploring} perform a large-scale quantitative study of the misogynistic language across the Manosphere on Reddit. 
They create nine lexicons of misogynistic terms to investigate how misogynistic language is used in 6M posts from Manosphere-related subreddits.
Jaki et al.~\cite{jaki2019online} study misogyny on the Incels.me forum, analyzing users' language and detecting misogyny instances, homophobia, and racism using a deep learning classifier that achieves up to 95\% accuracy. 

Furthermore, Ribeiro et al.~\cite{ribeiro2020pick} focus on the evolution of the broader Manosphere and perform a large-scale characterization of multiple Manosphere communities mainly on Reddit and six other Web forums associated with these communities.
They find that older Manosphere communities, such as Men's Rights Activists and Pick Up Artists, are becoming less popular and active. 
In comparison, newer communities like Incels and MGTOWs attract more attention. 
They also find a substantial migration of users from old communities to new ones, and that newer communities harbor more toxic and extreme ideologies.
In another study, Ribeiro et al.~\cite{manoel_bans} investigate whether platform migration of toxic online communities compromises content moderation.
To do this, they focus on two communities on Reddit, namely, /r/Incels and /r/The\_Donald, and use them to assess whether community-level moderation measures were effective in reducing the negative impact of toxic communities.
They conclude that a given platforms’ moderation measures may create even more radical communities on other platforms.
Instead, in our work we focus on the most extreme subgroup of the Manosphere, the Incel community, and we provide the first study of this community on YouTube, a platform where misogynistic ideologies, like Incel ideology, are relatively unstudied.  
We focus on analyzing the footprint of this community on YouTube aiming to quantify its growth over the last decade.
More importantly, we also investigate how the opaque nature of YouTube’s recommendation algorithm enables the discovery of Incel-related content by both random users of the platform and users who purposefully choose to see such content.

\descr{Harmful Activity on YouTube.}
YouTube's role in harmful activity has been studied mostly in the context of detection.
Agarwal et al.~\cite{agarwal2014focused} present a binary classifier trained with user and video features to detect videos promoting hate and extremism on YouTube, while Giannakopoulos et al.~\cite{giannakopoulos2010multimodal} develop a k-nearest classifier trained with video, audio, and textual features to detect violence on YouTube videos. Jiang et al.~\cite{jiang2019bias} investigate how channel partisanship and video misinformation affect comment moderation on YouTube, finding that comments are more likely to be moderated if the video channel is ideologically extreme.
Sureka et al.~\cite{sureka2010mining} use data mining and social network analysis techniques to discover hateful YouTube videos, while Ottoni et al.~\cite{ottoni2018analyzing} analyze video content and user comments on alt-right channels. 
Zannettou et al.~\cite{zannettou2018good} present a deep learning classifier for detecting videos that use manipulative techniques to increase their views, i.e., clickbait.
Papadamou et al.~\cite{papadamou2020disturbed}, and Tahir et al.~\cite{tahir2019BringingKidBacka} focus on detecting inappropriate videos targeting children on YouTube. Mariconti et al.~\cite{enrico2019cscw} build a classifier to predict, at upload time, whether or not a YouTube video will be ``raided'' by hateful users.

\descr{Calls for action.}
Additional studies point to the need for a better understanding of misogynistic content on YouTube.
Wotanis et al.~\cite{wotanis2014performing} show that more negative feedback is given to female than male YouTubers by analyzing hostile and sexist comments on the platform. D{\"o}ring et al.~\cite{doring2019male} build on this study by empirically investigating male dominance and sexism on YouTube, concluding that male YouTubers dominate YouTube, and that female content producers are prone to receiving more negative and hostile video comments.

To the best of our knowledge, our work is the first to provide a large-scale understanding and analysis of misogynistic content on YouTube generated by the Manosphere subgroups.  In particular, we investigate the role of YouTube's recommendation algorithm in disseminating Incel-related content on the platform.

\begin{table}[t!]
\setlength{\tabcolsep}{3pt}
\resizebox{\columnwidth}{!}{
\begin{tabular}{lrrrrrrr}
\toprule
\textbf{Subreddit} & \textbf{\#Videos} & \textbf{\#Users} & \textbf{\#Posts} & \textbf{Min. Date} & \textbf{Max. Date} & \textbf{\#Incel-related} & \textbf{\#Other}\\
 & & & & & & \textbf{Videos} & \textbf{Videos} \\
\midrule
Braincels       & 2,744 & 2,830,522 & 51,443  & 2017-10 & 2019-05 & 175 & 2,569\\
ForeverAlone    & 1,539 & 1,921,363 & 86,670    & 2010-09 & 2019-05 & 45 & 1,494\\
IncelTears      & 1,285 & 1,477,204 & 93,684  & 2017-05 & 2019-05 & 56 & 1,229\\
Incels        & 976   & 1,191,797 & 39,130  & 2014-01 & 2017-11 & 48 & 928\\
IncelsWithoutHate   & 223   & 163,820 & 7,141   & 2017-04 & 2019-05 & 16 & 207\\
ForeverAloneDating  & 92  & 153,039 & 27,460  & 2011-03 & 2019-05 & 0 & 92\\
askanincel      & 25  & 39,799  & 1,700   & 2018-11 & 2019-05 & 2 & 23\\
BlackPillScience  & 25  & 9,048   & 1,363   & 2018-03 & 2019-05 & 5 & 20\\
ForeverUnwanted   & 23  & 24,855  & 1,136   & 2016-02 & 2018-04 & 4 & 19\\
Incelselfies    & 17  & 60,988  & 7,057   & 2018-07 & 2019-05 & 1 & 16\\
Truecels      & 15  & 6,121   & 714   & 2015-12 & 2016-06 & 1 & 14\\
gymcels       & 5   & 1,430   & 296   & 2018-03 & 2019-04 & 2 & 3\\
MaleForeverAlone  & 3   & 6,306   & 831     & 2017-12 & 2018-06 & 0 & 3\\
foreveraloneteens   & 2   & 2,077   & 450   & 2011-11 & 2019-04 & 0 & 2\\
gaycel        & 1   & 117     & 43    & 2014-02 & 2018-10 & 0 & 1\\
SupportCel      & 1   & 6,095   & 474     & 2017-10 & 2019-01 & 0 & 1\\
Truefemcels     & 1   & 311     & 95    & 2018-09 & 2019-04 & 0 & 1\\
Foreveralonelondon  & 0   & 57    & 19    & 2013-01 & 2019-01 & 0 & 0\\
IncelDense      & 0   & 2,058   & 388     & 2018-06 & 2019-04 & 0 & 0\\
\midrule
\textbf{Total (Unique)}    & \textbf{6,977} & \textbf{7,897,007} & \textbf{320,094} & \textbf{-} & \textbf{-} & \textbf{290} & \textbf{6,162}\\ 
\bottomrule
\end{tabular}}
\caption{Overview of our Reddit dataset. We also include, for each subreddit, the number of videos from our Incel-derived labeled dataset.
The total number of videos reported in the individual subreddits differs from the unique videos collected since multiple videos have been shared in more than one subreddit.}
\label{tab:reddit_dataset_overview}
\end{table}

\descr{YouTube Recommendations.}
YouTube determines the ranks of the videos recommended to users based on various user engagement (e.g., user clicks, degree of engagement with recommended videos, etc.) and satisfaction metrics (e.g., likes, dislikes, etc.).
Aiming to increase the time that a user spends watching a particular video, the platform also considers various other user personalization factors, such as demographics, geolocation, or the watch history of the user~\cite{zhao2019recommending}.

Covington et al.~\cite{covington2016deep} describe YouTube's recommendation algorithm, using a deep candidate generation model to retrieve a small subset of videos from a large corpus and a deep ranking model to rank those videos based on their relevance to the user's activity. 
Zhao et al.~\cite{zhao2019recommending} propose a large-scale ranking system for YouTube recommendations. 
The proposed model ranks the candidate recommendations based on user engagement and satisfaction metrics.

Others focus on analyzing YouTube recommendations on specific topics.
Ribeiro et al.~\cite{ribeiro2019auditing} perform a large-scale audit of user radicalization on YouTube: they analyze videos from Intellectual Dark Web, Alt-lite, and  Alt-right channels, showing that they increasingly share the same user base.
They also analyze YouTube's recommendation algorithm finding that Alt-right channels can be reached from both Intellectual Dark Web and Alt-lite channels.
St{\"o}cker et al.~\cite{stocker2020riding} analyze the effect of extreme recommendations on YouTube, finding that YouTube's auto-play feature is problematic.
They conclude that preventing inappropriate personalized recommendations is technically infeasible due to the nature of the recommendation algorithm.
Finally,~\cite{hussein2020measuring} focus on measuring misinformation on YouTube and perform audit experiments considering five popular topics like 9/11 and chemtrail conspiracy theories to investigate whether personalization contributes to amplifying misinformation.
They audit three YouTube features: search results, Up-next video, and Top 5 video recommendations, finding a filter bubble effect~\cite{pariser2011filter} in the video recommendations section for almost all the topics they analyze. 
In contrast to the above studies, we focus on a different societal problem on YouTube. 
We explore the footprint of the Incel community, and we analyze the role of the recommendation algorithm in nudging users towards them. 
To the best of our knowledge, our work is the first to study the Incel community on YouTube and the role of YouTube's recommendation algorithm in the circulation of Incel-related content on the platform. 
We devise a methodology for annotating videos on the platform as Incel-related and using several tools, including text and graph analysis. 
We study the Incel community's footprint on YouTube and assess how YouTube’s recommendation algorithm behaves with respect to Incel-related videos.

\section{Dataset}
\label{sec:data_collection}
We now present our data collection and annotation process to identify Incel-related videos.

\subsection{Data Collection}
To collect Incel-related videos on YouTube, we look for YouTube links on Reddit, since recent work~\cite{ribeiro2020pick} highlighted that Incels are particularly active on Reddit.

We start by creating a list of subreddits that are relevant to Incels.
To do so, we inspect around $15$ posts on the Incel Wiki~\cite{incelswiki_2019} looking for references to subreddits and compile a list\footnote{Available at \url{https://bit.ly/incel-related-subreddits-list}.} of $19$ Incel-related subreddits.
This list also includes a set of communities relevant to Incel ideology (even possibly anti-incel like /r/Inceltears) to capture a broader collection of relevant videos. 

\begin{figure*}[t!]
\centering
\includegraphics[width=\linewidth]{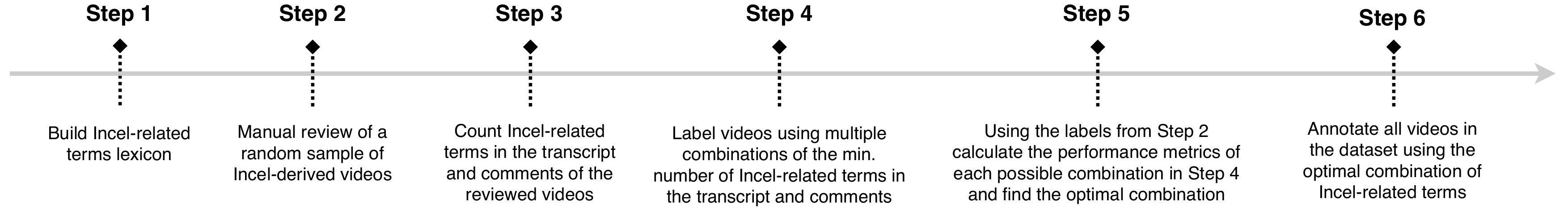}
\caption{Overview of our video annotation methodology.}
\label{fig:video_annotation_methodology}
\end{figure*}

We collect all submissions and comments made between June 1, 2005, and April 30, 2019, on the 19 Incel-related subreddits using the Reddit monthly dumps from Pushshift~\cite{baumgartner2020pushshift}. 
We parse them to gather links to YouTube videos, extracting 5M posts, including 6.5K unique links to YouTube videos that are still online and have a transcript available by YouTube to download. Next, we collect the metadata of each YouTube video using the YouTube Data API~\cite{youtubedataapi_2019}.
Specifically, we collect: 1) transcript; 2) title and description; 3) a set of tags defined by the uploader; 4) video statistics such as the number of views, likes, etc.; and 5) the top 1K comments, defined by YouTube's relevance metric, and their replies. Throughout the rest of this paper, we refer to this set of videos, which is derived from Incel-related subreddits, as {\bf\em ``Incel-derived''} videos.

Table~\ref{tab:reddit_dataset_overview} reports the total number of users, posts, linked YouTube videos, and the period of available information for each subreddit.
Although recently created, /r/Braincels has the largest number of posts and YouTube videos. Also, even though it was banned in November 2017 for inciting violence against women~\cite{incelssubredditbanned_2017}, /r/Incels is fourth in terms of YouTube videos shared. Lastly, note that most of the subreddits in our sample were created between 2015 and 2018, which already suggests a trend of increasing popularity for the Incel community.

\descr{Control Set.} 
\label{subsec:control_videos}
We also collect a dataset of random videos and use it as a control to capture more general trends on YouTube videos shared on Reddit as the Incel-derived set includes only videos posted on Incel communities on Reddit. 
To collect Control videos, we parse all submissions and comments made on Reddit between June 1, 2005, and April 30, 2019, using the Reddit monthly dumps from Pushshift, and we gather all links to YouTube videos.  
From them, we randomly select 5,793 links shared in 2,154 subreddits\footnote{See \url{https://bit.ly/incels-control-videos-subreddits} for the list of subreddits and the number of control videos shared in each subreddit.} for which we collect their metadata using the YouTube Data API. 

We choose to use a randomly selected set of videos shared on Reddit as our Control set for a more fair comparison since our Incel-derived set also includes videos shared on this platform.
We collect random videos instead of videos relevant to another sensitive topic because this allows us to study the amount of Incel-related content that can generally be found on YouTube.
At the same time, videos relevant to another sensitive topic or community (e.g., MGTOW) may have strong similarities with Incel-related videos, hence they may not be able to capture more general trends on YouTube.

\begin{table}[t!]
\centering
\small
\resizebox{\columnwidth}{!}{
\begin{tabular}{rr|rrrr}
\toprule
\multicolumn{2}{c|}{\bf \#Incel-related Terms} & & & &\\
  {\bf\em in Transcript} & {\bf\em in Comments} &
{\textbf{Accuracy}} &
{\textbf{Precision}} &
{\textbf{Recall}} &
{\textbf{F1 Score}} \\ 
\midrule
$\geq$0          & $\geq$7          & 0.81          & 0.77          & 0.80          & 0.78          \\
$\geq$1          & $\geq$1          & 0.82          & 0.78          & 0.82          & 0.79          \\
$\geq$0          & $\geq$3          & 0.79          & 0.79          & 0.79          & 0.79          \\
$\geq$1          & $\geq$2          & 0.83          & 0.78          & 0.83          & 0.79          \\
\textbf{$\geq$1} & \textbf{$\geq$3} & \textbf{0.83} & \textbf{0.79} & \textbf{0.83} & \textbf{0.79} \\
\bottomrule
\end{tabular}%
}
\caption{Performance metrics of the top combinations of the number of Incel-related terms in a video's transcript and comments.}
\label{tab:thresholds_performance_metrics}
\end{table}

\subsection{Video Annotation}\label{subsec:video_annotation}
The analysis of Incel-related content on YouTube differs from analyzing other types of inappropriate content on the platform.
So far, there is no prior study exploring the main themes involved in videos that Incels find of interest. 
This renders the task of annotating the actual video rather cumbersome. 
Besides, annotating the video footage does not by itself allow us to study the footprint of the Incel community on YouTube effectively. 
When it comes to this community, it is not only the video's content that may be relevant. Rather, the language that the community members use in their videos or comments for or against their views is also of interest. 
For example, there are videos featuring women talking about feminism, which are heavily commented on by Incels.

\descr{Building a Lexicon.} 
To capture the variety of aspects of the problem, we devise an annotation methodology based on a lexicon of terms that are routinely used by members of the Incel community and use it to annotate the videos in our dataset. 
Figure~\ref{fig:video_annotation_methodology} depicts the individual steps that we follow in the devised video annotation methodology.

To create the lexicon (Step 1 in Figure~\ref{fig:video_annotation_methodology}), we first crawl the ``glossary'' available on the Incels Wiki page~\cite{incelglossary_2019}, gathering 395 terms. 
Since the glossary includes several words that can also be regarded as general-purpose (e.g., fuel, hole, legit, etc.), we employ three human annotators to determine whether each term is specific to the Incel community.

We note that all annotators label all the 395 terms of the glossary. 
The three annotators are authors of this paper and they are familiar with scholarly articles on the Incel community and the Manosphere in general.
Before the annotation task, a discussion took place to frame the task and the annotators were told to consider a term relevant only if it expresses hate, misogyny, or is directly associated with Incel ideology.
For example, the phrase ``Beta male'' or any Incel-related incident (e.g., ``supreme gentleman,''  an indirect reference to the Isla Vista killer Elliot Rodgers~\cite{elliotrodgers_2014}).
We note that, during the labeling, the annotators had no discussion or communication whatsoever about the task at hand.

We then create our lexicon by only considering the terms annotated as relevant, based on all the annotators' majority agreement, which yields a 200 Incel-related term dictionary\footnote{See \url{https://bit.ly/incel-related-terms-lexicon} for the final lexicon with all the relevant terms.}. 
We also compute the Fleiss' Kappa Score~\cite{fleiss1971measuring} to assess the agreement between the annotators, finding it to be $0.69$, which is considered ``substantial'' agreement~\cite{landis1977MeasurementObserverAgreementa}. 

\descr{Labeling.} Next, we use the lexicon to label the videos in our dataset. We look for these terms in the transcript, title, tags, and comments of our dataset videos. Most matches are from the transcript and the videos' comments; thus, we decide to use these to determine whether a video is Incel-related.
To select the minimum number of Incel-related terms that transcripts and comments should contain to be labeled as ``Incel-related,'' we devise the following methodology:

\begin{enumerate}
  \item We randomly select 1K videos from the Incel-derived set, which the first author of this paper manually annotates as ``Incel-related'' or ``Other'' by watching them and looking at the metadata.
   Note that Incel-related videos are a subset of Incel-derived ones (Step 2 in Figure~\ref{fig:video_annotation_methodology}).

  \item We count the number of Incel-related terms in the transcript and the annotated videos' comments (Step 3 in Figure~\ref{fig:video_annotation_methodology}).

  \item For each possible combination of the minimum number of Incel-related terms in the transcript and the comments, we label each video as Incel-related or not, and calculate the accuracy, precision, recall, and F1 score based on the labels assigned to the videos during the manual annotation (Steps 4 and 5 in Figure~\ref{fig:video_annotation_methodology}).
\end{enumerate}
Table~\ref{tab:thresholds_performance_metrics} shows the performance metrics for the top five combinations of the number of Incel-related terms in the transcript and the comments. 
We pick the one yielding the best F1 score (to balance between false positives and false negatives), which is reached if we label a video as Incel-related when there is at least one Incel-related term in the transcript and at least three in the comments. 
Using this rule, we annotate all the videos in our dataset (Steps 5 and 6 in Figure~\ref{fig:video_annotation_methodology}).

Table~\ref{tab:reddit_dataset_overview} reports the label statistics of the Incel-derived videos per subreddit.
Our final labeled dataset includes $290$ Incel-related and $6,162$ Other videos in the Incel-derived set, and $66$ Incel-related and $5,727$ Other videos in the Control set.

\begin{figure}[t!]
\centering
\subfigure[]{\includegraphics[width=\columnwidth]{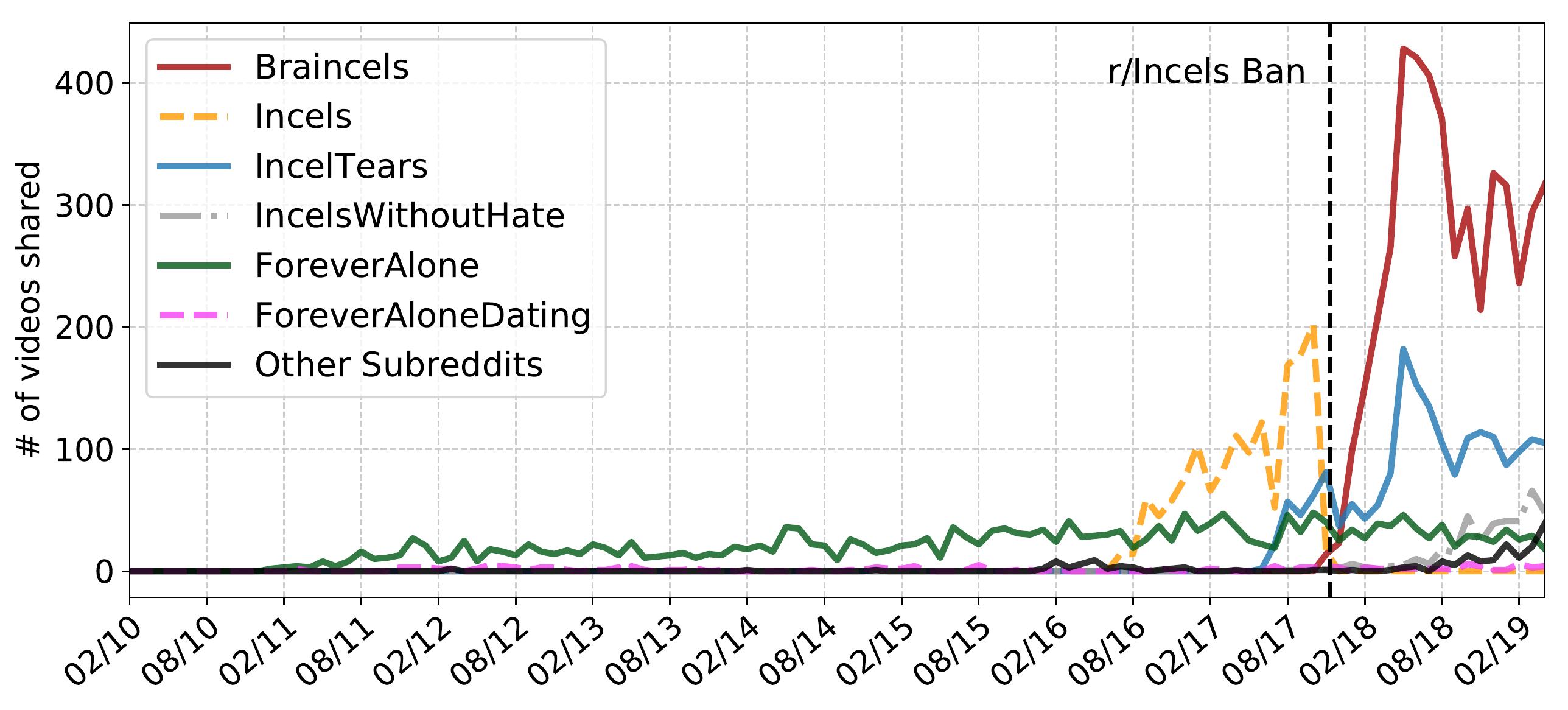}\label{fig:subreddits_youtube_links_shared_per_year}}
\subfigure[]{\includegraphics[width=\columnwidth]{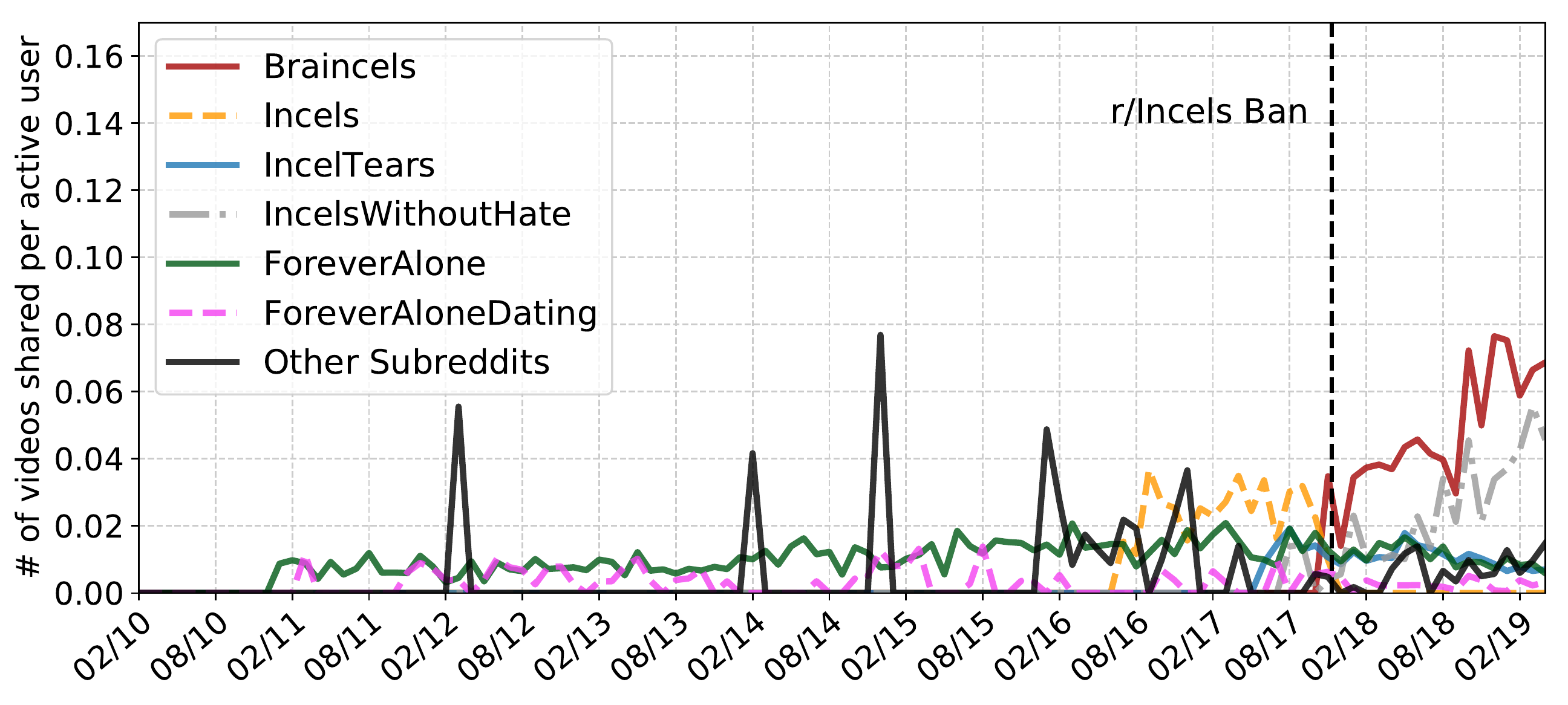}\label{fig:normalized_subreddits_youtube_links_shared_per_year}}
\caption{Temporal evolution of the number of YouTube videos shared on each subreddit per month. (a) depicts the absolute number of videos shared in each subreddit and (b) depicts the normalized number of videos shared per active user of each subreddit.
We annotate both figures with the date when Reddit decided to ban the /r/Incels subreddit.}
\label{fig:subreddits_youtube_links_shared_plot}
\end{figure}

\subsection{Ethics} 
Overall, we follow standard ethical guidelines~\cite{dittrich2012menlo,rivers2014ethical} regarding information research and the use of shared measurement data.
In this work, we only collect and process publicly available data, make no attempt to de-anonymize users, and our data collection does not violate the terms of use of the APIs we employ. 
More precisely, we ensure compliance with GDPR's ``Right of Access''~\cite{gdpr2018rightaccess} and ``Right to be Forgotten''~\cite{gdpr2018rightforgotten} principles. 
For the former, we give users the right to obtain a copy of any data that we maintain about them for the purposes of this research, while for the former we ensure that we delete and not share with any unauthorized party any information that has been deleted from the public repositories from which we obtain our data.
We also note that we do not share with anyone any sensitive personal data, such as the actual content of the comments that we analyze or the usernames of the commenting users.
Instead, we make publicly available for reproducibility and research purposes all the metadata of the collected and annotated videos\footnote{\url{https://zenodo.org/record/4557039}} that do not include any personal data, as well as the unique identifiers of the comments that we analyze while ensuring that we abide by GDPR's ``Right to be Forgotten''~\cite{gdpr2018rightforgotten} principle.

Furthermore, our video annotation methodology abides by the ethical guidelines defined by the Association of Internet Researchers (AoIR) for the protection of researchers~\cite{aoir2019ethicalguidelines}.
Note that in our video annotation methodology we do not engage any human subjects other than the three authors of this paper.
Since the annotators are authors of this paper, we do not take into consideration harmful effects on random human annotators due to inappropriate content.
However, we still consider the effect of the content that we study on the authors and especially the student authors. 
We address this with continuous monitoring and open discussions with members of the research team, as
well as by properly applying best practices from the psychological and social scientific literature on the topic, e.g.,~\cite{beale2004impact,bashir2018doing,blagden2010challenge}. 
One of the primary goals is to minimize the risk that the researchers become de-sensitized with respect to such content.
Finally, we believe that studying misogynistic and hateful communities in depth is bound to be beneficial for society at large, as well as for victims of such abuse.

\begin{figure}[t!]
\centering
\includegraphics[width=\columnwidth]{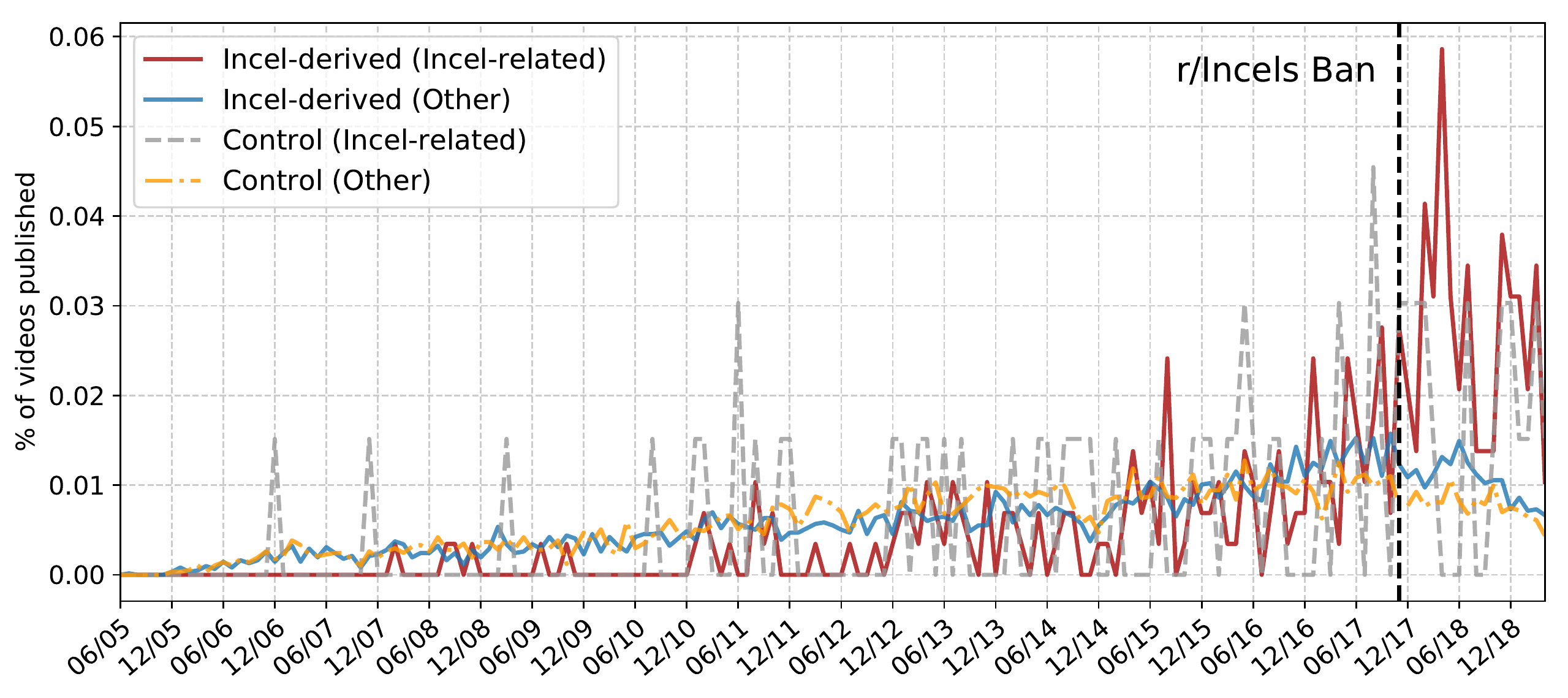}
\caption{Percentage of videos published per month for both Incel-derived and Control videos. We also depict the date when Reddit decided to ban the /r/Incels subreddit.}
\label{fig:videos_published_cumulative}
\end{figure}

\section{RQ1: Evolution of Incel community on YouTube}
\label{sec:temporal_analysis}
This section explores how the Incel communities on YouTube and Reddit have evolved in terms of videos and comments posted.

\subsection{Videos}
We start by studying the ``evolution'' of the Incel communities concerning the number of videos they share.
First, we look at the frequency with which YouTube videos are shared on various Incel-related subreddits per month; see Figure~\ref{fig:subreddits_youtube_links_shared_plot}.
Figure~\ref{fig:subreddits_youtube_links_shared_per_year} shows the absolute number of videos shared in each Incel-related subreddit per month, while Figure~\ref{fig:normalized_subreddits_youtube_links_shared_per_year} shows the number of videos shared in each subreddit per active user of each community.
After June 2016, we observe that Incel-related subreddits users start linking to YouTube videos more frequently and more in 2018. 
This trend is more pronounced on /r/Braincels in both the absolute number of videos shared and the number of videos shared per active user; see /r/Braincels in Figure~\ref{fig:subreddits_youtube_links_shared_per_year} and Figure~\ref{fig:normalized_subreddits_youtube_links_shared_per_year}.
This indicates that the use of YouTube to spread Incel ideology is increasing.
Note that the sharp drop of /r/Incels activity is due to Reddit's decision to ban this subreddit for inciting violence against women in November 2017~\cite{incelssubredditbanned_2017} (see annotation in Figure~\ref{fig:subreddits_youtube_links_shared_per_year} and Figure~\ref{fig:normalized_subreddits_youtube_links_shared_per_year}).
However, the sharp increase of /r/Braincels activity after this period questions the efficacy of Reddit's decision to ban /r/Incels, and it can be considered as evidence that the ban was ineffective in terms of suppressing the activity of the Incel community on the platform.
It also worths noting that Reddit decided to ban /r/Braincels in September 2019~\cite{theverge2019raincelsban}.

\begin{figure}[t!]
\centering
\includegraphics[width=\columnwidth]{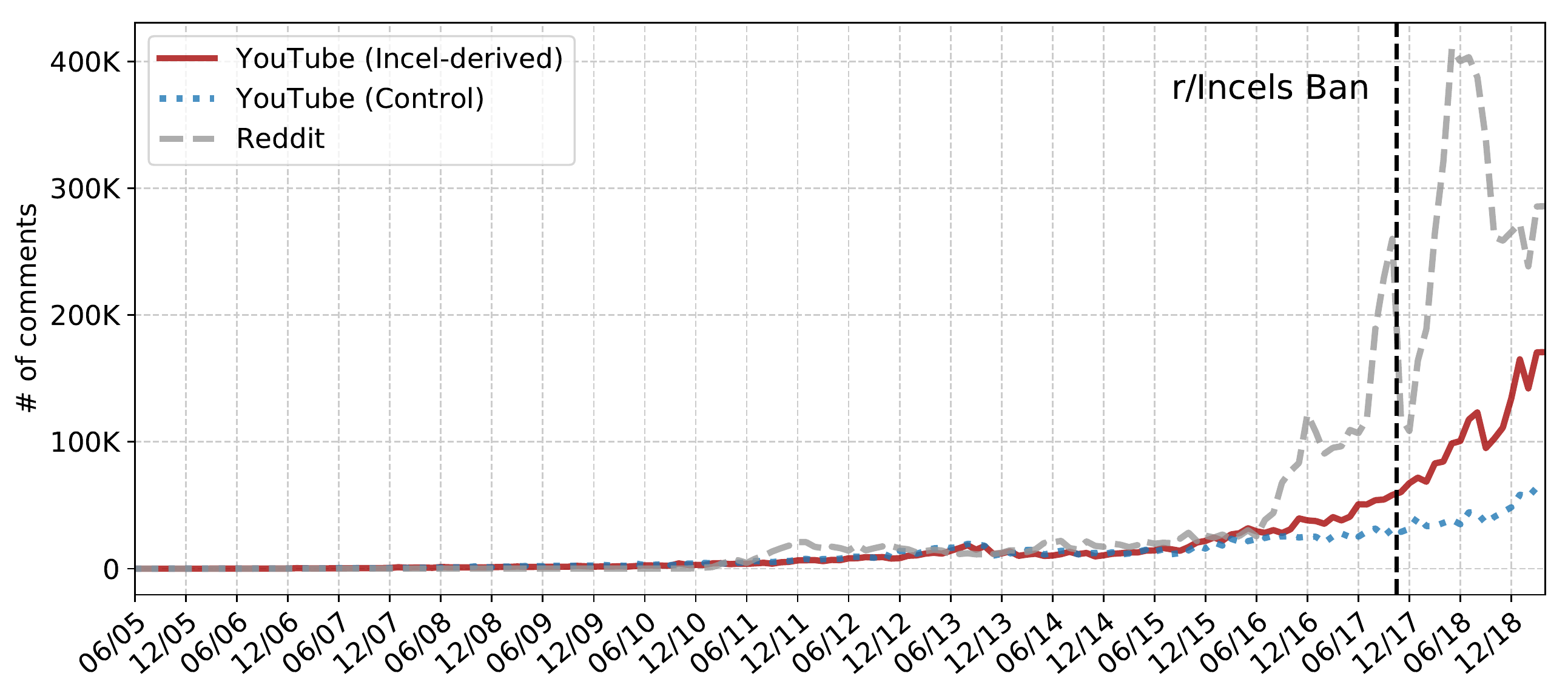}
\caption{Temporal evolution of the number of comments per month. We also depict the date when Reddit decided to ban the /r/Incels subreddit.}
\label{fig:number_of_comments_per_date_plot}
\end{figure}

\begin{figure}[t!]
\centering
\includegraphics[width=\columnwidth]{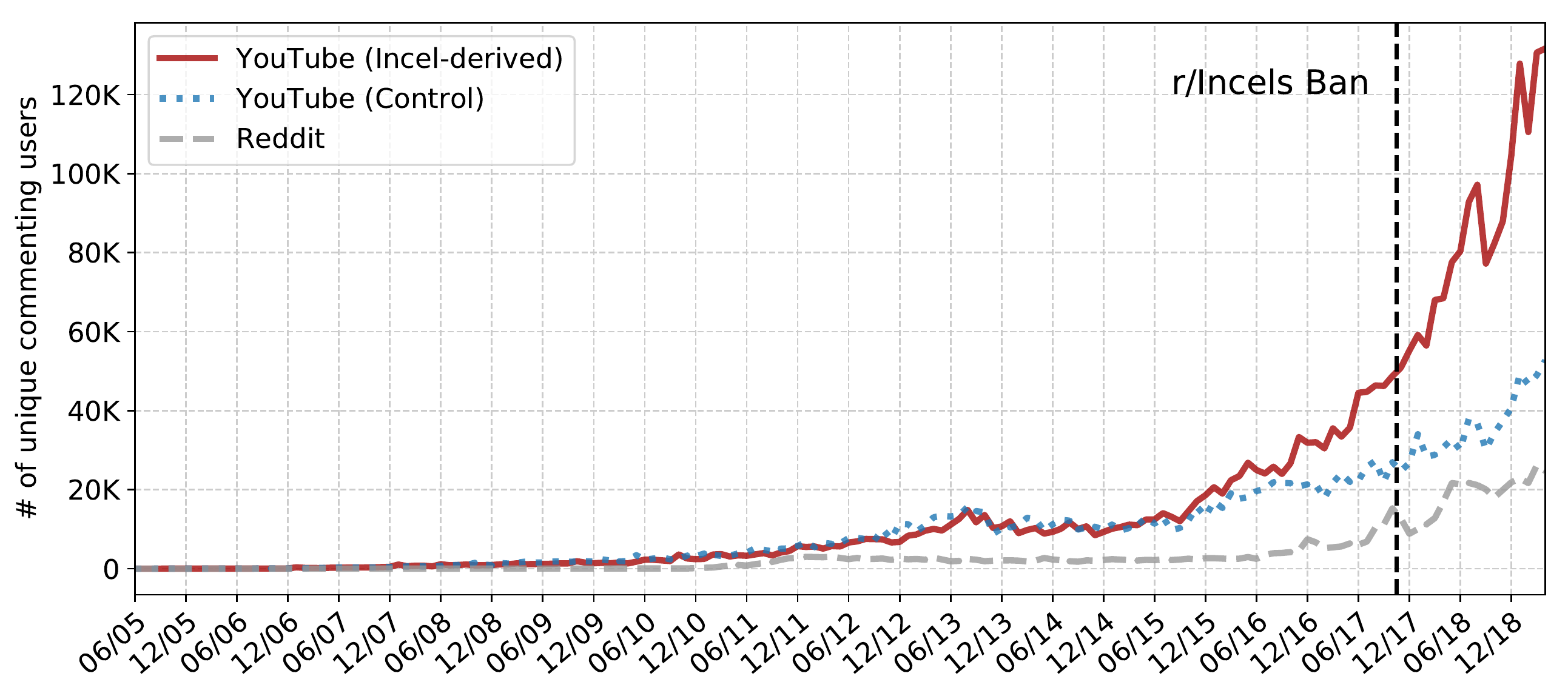}
\caption{Temporal evolution of the number of unique commenting users per month. We also depict the date when Reddit decided to ban the /r/Incels subreddit.}
\label{fig:unique_commenting_users_per_date_plot}
\end{figure}

In Figure~\ref{fig:videos_published_cumulative}, we plot the percentage of videos published per month for both Incel-derived and Control videos, while we also depict the date when Reddit decided to ban the /r/Incels subreddit.
While the increase in the number of Other videos remains relatively constant over the years for both sets of videos, this is not the case for Incel-related ones, as $81\%$ and $64\%$ of them in the Incel-derived and Control sets, respectively, published after December 2014.
Overall, there is a steady increase in Incel activity, especially after 2016, which is particularly worrisome as we have several examples of users who were radicalized online and have gone to undertake deadly attacks~\cite{cecco_2019}.
An even higher increase in Incel-activity is also observed after the ban of the /r/Incels subreddit.

\subsection{Comments}
Next, we study the commenting activity on both Reddit and YouTube.
Figure~\ref{fig:number_of_comments_per_date_plot} shows the number of comments posted per month for both YouTube Incel-derived and Control videos, and Reddit.
Activity on both platforms starts to markedly increase after 2016, and more after the ban of /r/Incels in November 2017, with Reddit and YouTube Incel-derived videos having substantially more comments than the Control videos.
Once again, the sharp increase in the commenting activity over the last few years signals an increase in the Incel user base's size.

To further analyze this trend, we look at the number of {\em unique} commenting users per month on both platforms; see Figure~\ref{fig:unique_commenting_users_per_date_plot}.
On Reddit, we observe that the number of unique users remains steady over the years, increasing from 10K in August 2017 to 25K in April 2019. 
This is mainly because most of the subreddits in our dataset ($58\%$) were created after 2016.
On the other hand, for the Incel-derived videos on YouTube, there is a substantial increase from 30K in February 2017 to 132K in April 2019. 
We also observe an increase of the Control videos' unique commenting users (from 18K in February 2017 to 53K in April 2019).
However, the increase is not as sharp as that of the Incel-derived videos; $483\%$ vs. $1,040\%$ increase in the average unique commenting users per month after January 2017 in Control and Incel-derived videos, respectively.

\begin{figure}[t!]
\centering
\includegraphics[width=\columnwidth]{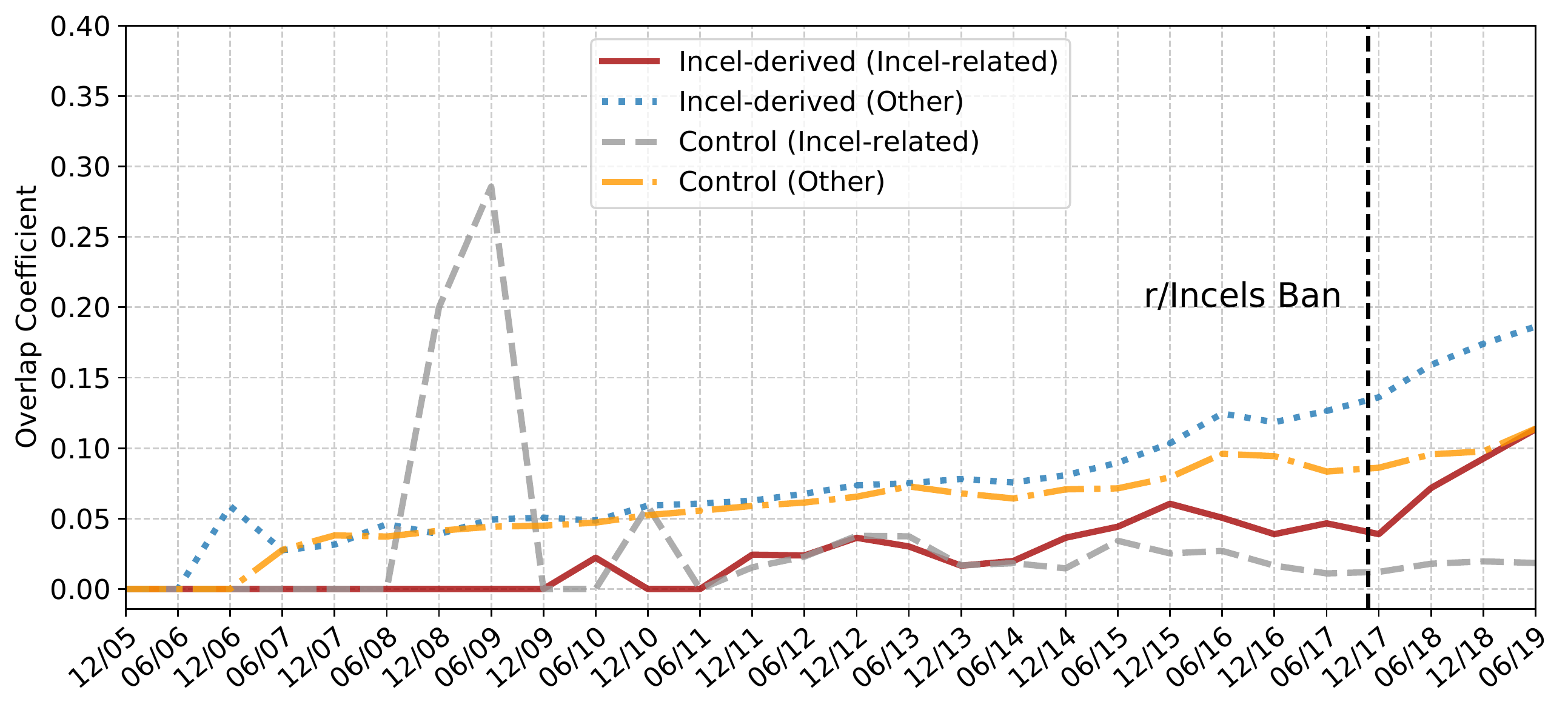}
\caption{Self-similarity of commenting users in adjacent months for both Incel-derived and Control videos. We also depict the date when Reddit decided to ban the /r/Incels subreddit.}
\label{fig:overlap_videos_users_similarity_per_year_plot}
\end{figure}

To assess whether the sharp increase in unique commenting users of the Incel-derived and Control videos after 2017 is due to the increased interest by random users or to an increased interest in those videos and their discussions by the same users over the years, we use the Overlap Coefficient similarity metric~\cite{vijaymeena2016survey}; it measures user retention over time for the videos in our dataset. 
Specifically, we calculate, for each month, the similarity of commenting users with those doing so {\em the month before}, for both Incel-related and Other videos in the Incel-derived and Control sets. 
Note that if the set of commenting users of a specific month is a subset of the previous month’s commenting users or the converse, the overlap coefficient is equal to 1.
The results of this calculation are shown in Figure~\ref{fig:overlap_videos_users_similarity_per_year_plot}, in which we again depict the date when Reddit decided to ban /r/Incels.
Interestingly, for the Incel-related videos of the Incel-derived set, we find a sharp growth in user retention right after the ban of the /r/Incels subreddit in November 2017, while this is not the case for the Incel-related videos of the Control set.
For the Incel-related videos of the Control set, we observe a more steady increase in user retention over time.
Once again, this might be related to the increased popularity of the Incel communities and might indicate that the ban of /r/Incels energized the community and made participants more persistent.
Also, the higher user retention of Other videos in both sets is likely due to the much higher proportion of Other videos in each set.

Last, we observe a spike in user retention for the Incel-related videos of the Control set during 2009.
However, after checking the publication dates of these videos in our dataset, we only find three Incel-related videos in the Control set uploaded before July 2009. 
Hence, it might be the case that the same users repeatedly commented on those videos during 2008 and 2009.
At the same time, no other Incel-related videos in the Control was uploaded between July 2009 and July 2010, hence the drop in user retention after July 2009.

\section{RQ2: Does YouTube's recommendation algorithm steer users towards Incel-related videos?} 
\label{sec:recommendation_analysis}
Next, we present an analysis of how YouTube's recommendation algorithm behaves with respect to Incel-related videos.
More specifically, 1) we investigate how likely it is for YouTube to recommend an Incel-related video;
2) we simulate the behavior of a user who views videos based on the recommendations by performing random walks on YouTube's recommendation graph to measure the probability of such a user discovering Incel-related content; and
3) we investigate whether the frequency with which Incel-related videos are recommended increase for users who choose to see the content.

\begin{table}[t!]
\centering
\small
\begin{tabular}{lrr}
\toprule
\textbf{Recommendation Graph} & \textbf{Incel-related} & \textbf{Other} \\
\midrule
Incel-derived & 1,074 ($2.9\%$) & 36,673 ($97.1\%$) \\
Control & 428 ($1.5\%$) & 28,866 ($98.5\%$) \\
\bottomrule
\end{tabular}%
\caption{Number of Incel-related and Other videos in each recommendation graph.}
\label{tab:videos_in_recommendations_rounds}
\end{table}

\subsection{Recommendation Graph Analysis}
To build the recommendation graphs used for our analysis, we use functionality provided by the YouTube Data API. 
For each video in the Incel-derived and Control sets, we collect the top 10 recommended videos associated with it.
Note that the use of the YouTube Data API is associated with a specific account only for authentication to the API, and that the API does not maintain a watch history nor any cookies.
Thus, our data collection does not capture how specific account features or the viewing history affect personalized recommendations.  
Instead, the YouTube Data API allows us to collect recommendations provided by YouTube's recommendation algorithm based on video item-to-item similarity, as well as general user engagement and satisfaction metrics~\cite{zhao2019recommending}. 
The collected recommendations are similar to the recommendations presented to a non-logged-in user who watches videos on YouTube.
We collect the recommendations for the Incel-derived videos between September 20 and October 4, 2019, and the Control between October 15 and November 1, 2019. 
To annotate the collected videos, we follow the same approach described in Section~\ref{subsec:video_annotation}. 
Since our video annotation is based on the videos' transcripts, we only consider the videos that have one when building our recommendations graphs.

Next, we build a directed graph for each set of recommendations, where nodes are videos (either our 
dataset videos or their recommendations), and edges between nodes indicate the recommendations between all videos (up to ten).
For instance, if \textit{video2} is recommended via \textit{video1}, then we add an edge from \textit{video1} to \textit{video2}. 
Throughout the rest of this paper, we refer to each set of videos' collected recommendations as separate \emph{recommendation graphs}.

First, we investigate the prevalence of Incel-related videos in each recommendation graph.
Table~\ref{tab:videos_in_recommendations_rounds} reports the number of Incel-related and Other videos in each graph.
For the Incel-derived graph, we find 36,7K ($97.1\%$) Other and 1K ($2.9\%$) Incel-related videos, while in the Control graph, we find 28,9K ($98.5\%$) Other and 428 ($1.5\%$) Incel-related videos. 
These findings highlight that despite the proportion of Incel-related video recommendations in the Control graph being smaller, there is still a non-negligible amount recommended to users. 
Also, note that we reject the null hypothesis that the differences between the two graphs are due to chance via the Fisher's exact test ($p<0.001$)~\cite{fisher1922interpretation}.

\begin{table}[t!]
\centering
\small
\resizebox{\columnwidth}{!}{
\begin{tabular}{llrr}
\toprule
\textbf{Source} & \textbf{Destination} & \textbf{Incel-derived} & \textbf{Control} \\ 
\midrule
Incel-related   & Incel-related & 889 ($0.8\%$)       & 89 ($0.2\%$) \\
Incel-related   & Other       & 3632 ($3.2\%$)    & 773 ($1.4\%$) \\
Other         & Other       & 104,706 ($93.2\%$)    & 54,787 ($97.0\%$) \\
Other         & Incel-related & 3,160 ($2.8\%$)     & 842 ($1.5\%$) \\ 
\bottomrule
\end{tabular}%
}
\caption{Number of transitions between Incel-related and Other videos in each recommendation graph.}
\label{tab:graph_transitions_in_recommendation_rounds}
\end{table}

\begin{figure*}[t!]
\centering
\subfigure[]{\includegraphics[width=.7\columnwidth]{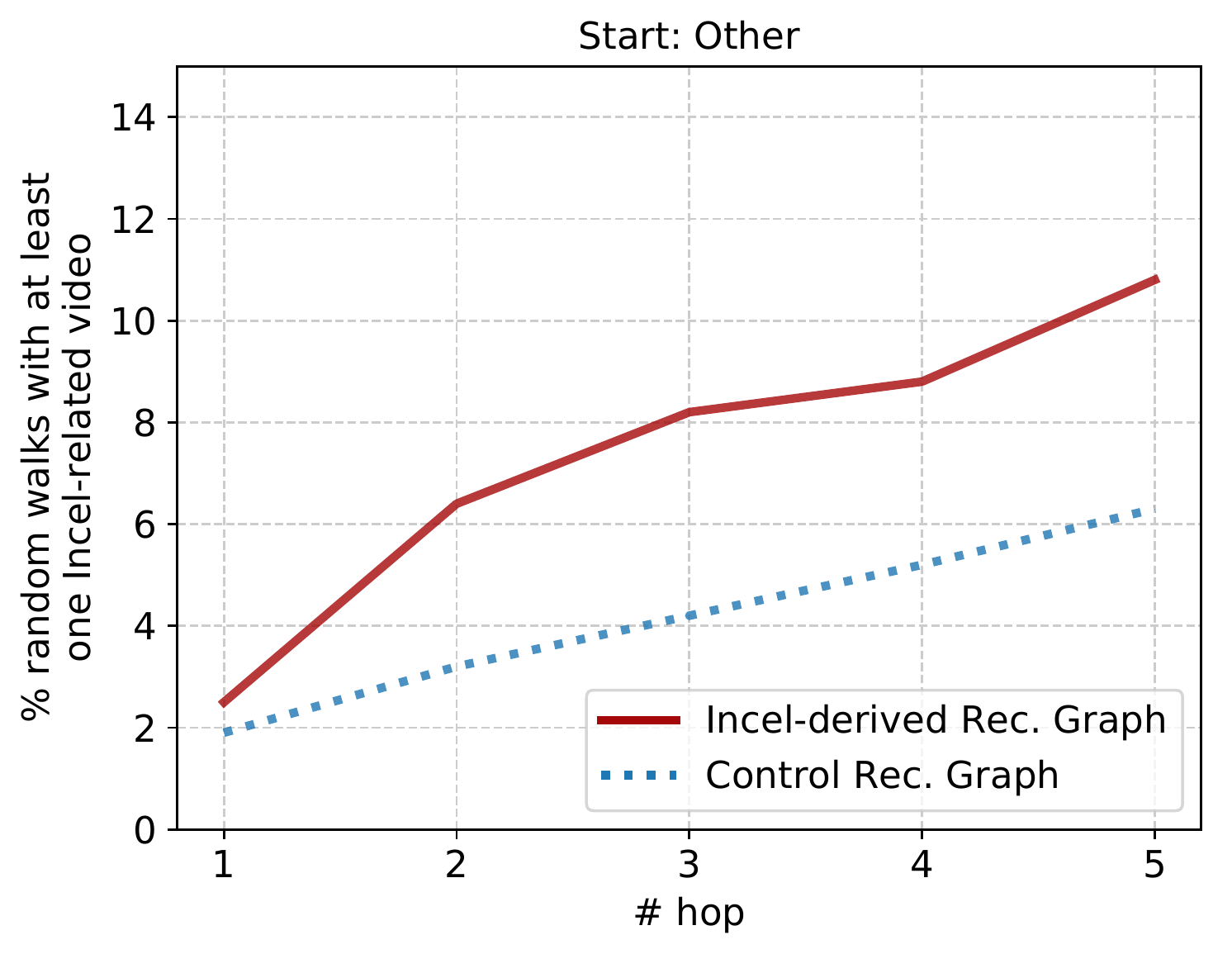}\label{fig:at_least_one_random_walks_plot_started_from_irrelevant}}
\subfigure[]{\includegraphics[width=.7\columnwidth]{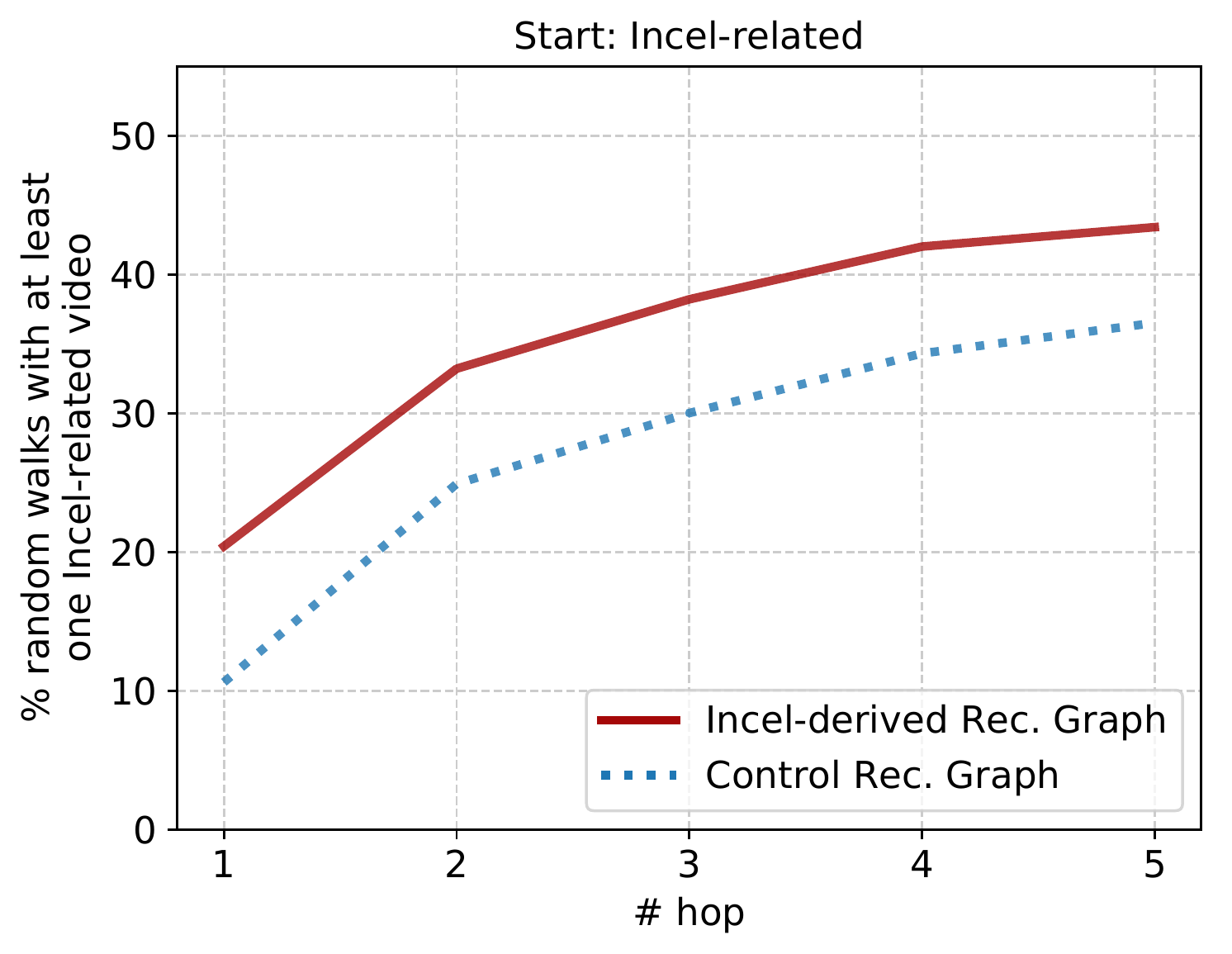}\label{fig:at_least_one_random_walks_plot_started_from_incels_related}}
\caption{Percentage of random walks where the random walker encounters at least one Incel-related video for both starting scenarios. Note that the random walker selects, at each hop, the next video to watch at random.}
\label{fig:random_walks_plots_at_least_one}
\end{figure*}

\begin{figure*}[t!]
\centering
\subfigure[]{\includegraphics[width=.7\columnwidth]{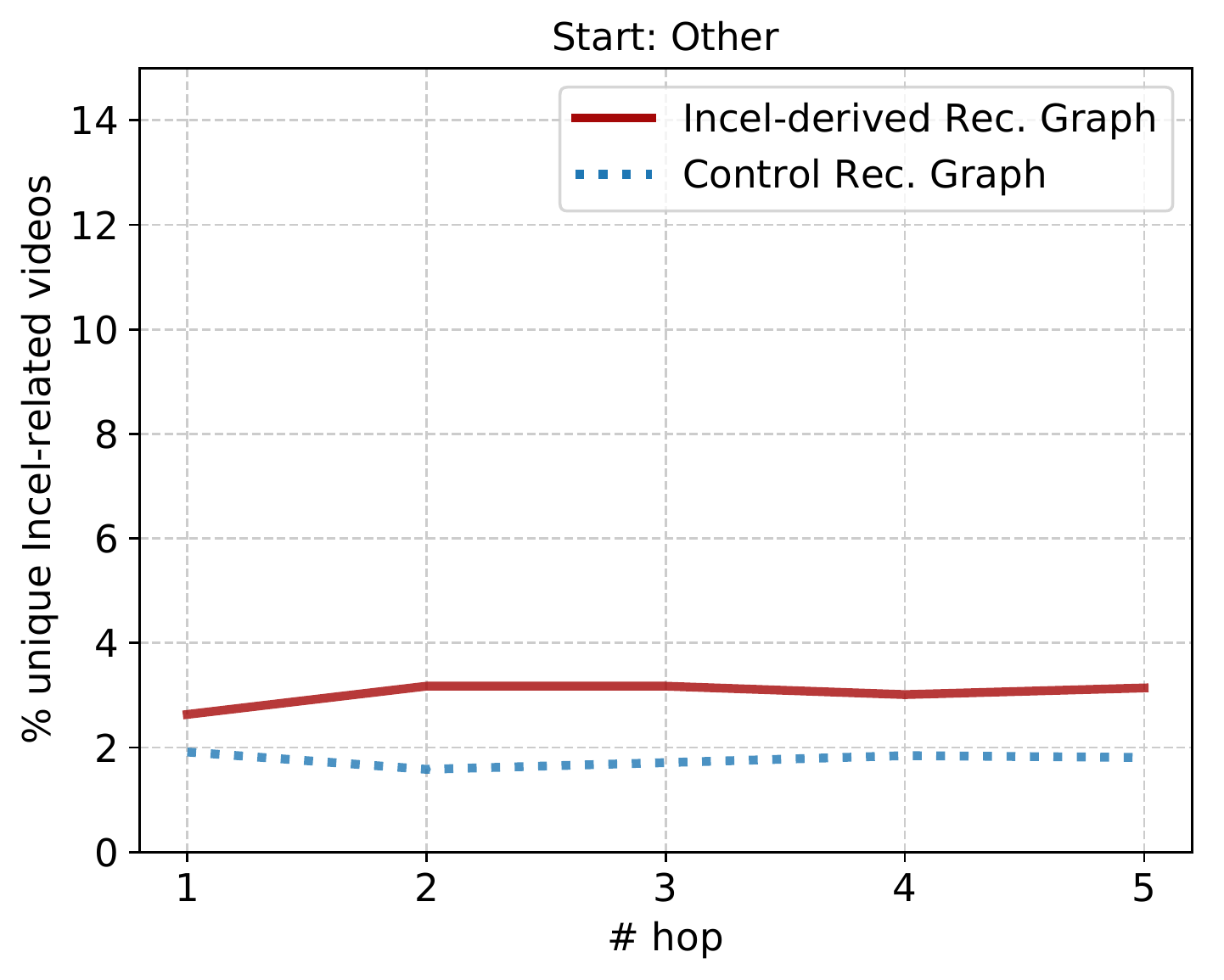}\label{fig:random_walks_plot_started_from_irrelevant}}
\subfigure[]{\includegraphics[width=.7\columnwidth]{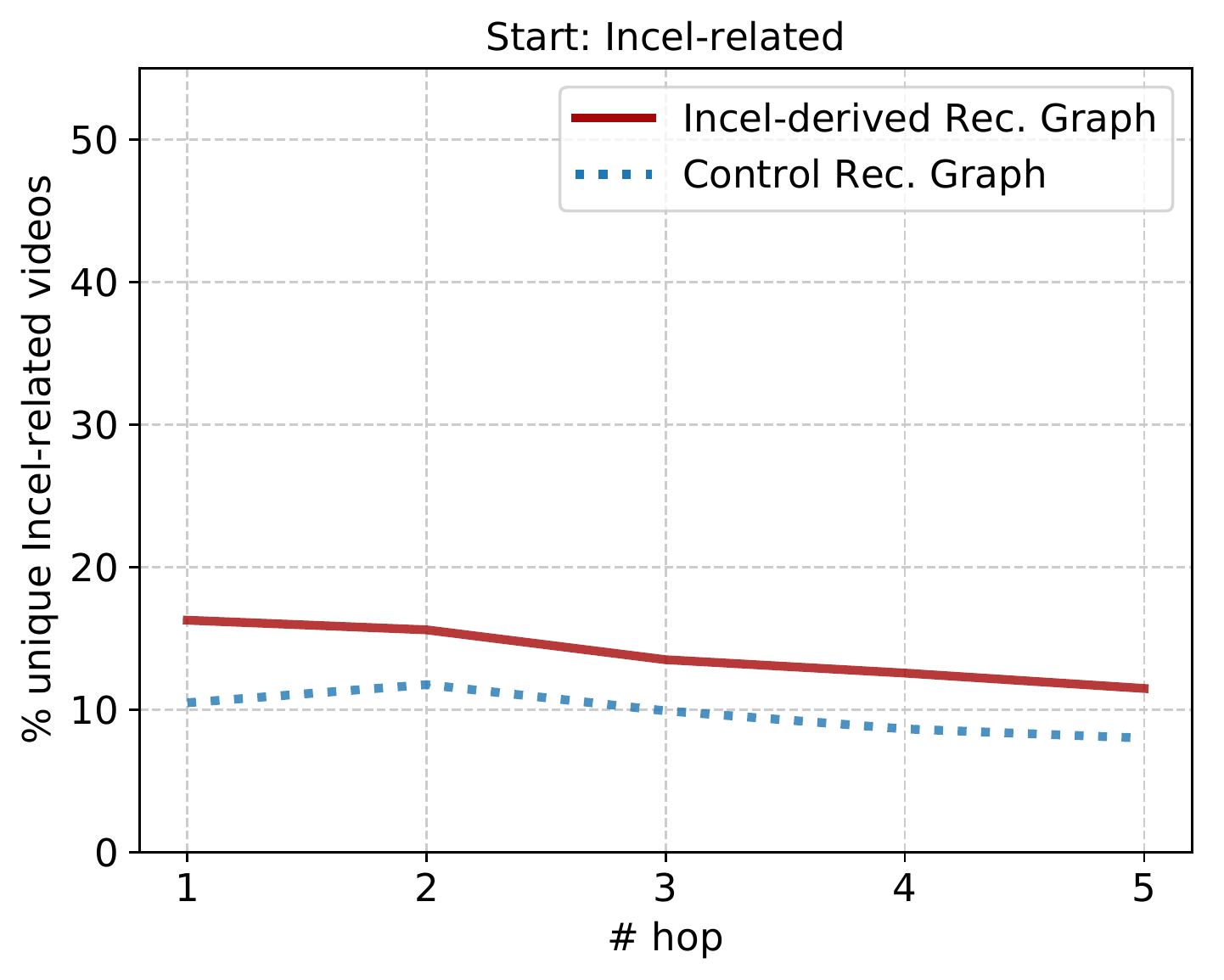}
\label{fig:random_walks_plot_started_from_incels_related}}
\caption{Percentage of Incel-related videos across all unique videos that the random walk encounters at hop $k$ for both starting scenarios. Note that the random walker selects, at each hop, the next video to watch at random.}
\label{fig:random_walks_plots}
\end{figure*}

\descr{How likely is it for YouTube to recommend an Incel-related Video?}
Next, to understand how frequently YouTube recommends an Incel-related video, we study the interplay between the Incel-related and Other videos in each recommendation graph. For each video, we calculate the out-degree in terms of Incel-related and Other labeled nodes.
We can then count the number of \textit{transitions} the graph makes between differently labeled nodes.
Table~\ref{tab:graph_transitions_in_recommendation_rounds} reports the percentage of each transition between the different types of videos for both graphs.
Perhaps unsurprisingly, most of the transitions, $93.2\%$ and $97.0\%$, respectively, in the Incel-derived and Control recommendation graphs are between Other videos, but this is mainly because of the large number of Other videos in each graph. We also find a high percentage of transitions between Other and Incel-related videos.
When a user watches an Other video, if they randomly follow one of the top ten recommended videos, there is a $2.9\%$ and $1.5\%$ probability in the Incel-derived and Control graphs, respectively, that they will end up at an Incel-related video.
Interestingly, in both graphs, Incel-related videos are more often recommended by Other videos than by Incel-related videos.
On the one hand, this might be due to the larger number of Other videos compared to Incel-related videos in both recommendation graphs.
On the other hand, this may indicate that YouTube's recommendation algorithm cannot discern Incel-related videos, which are likely misogynistic.

\begin{table*}[t!]
 \small
\centering
\begin{tabular}{>{\raggedleft\arraybackslash}p{1cm}
>{\raggedleft\arraybackslash}p{2.8cm}
>{\raggedleft\arraybackslash}p{2.8cm}
>{\raggedleft\arraybackslash}p{2.8cm}
>{\raggedleft\arraybackslash}p{2.8cm}}
\toprule
 & \multicolumn{2}{c}{\textbf{Incel-derived Recommendation Graph}} & \multicolumn{2}{c}{\textbf{Control Recommendation Graph}} \\
\midrule
\textbf{\#hop (M)} & \textbf{In next 5-M hops, $\geq$1 Incel-related} & \textbf{In next hop, 1 Incel-related} & \textbf{In next 5-M hops, $\geq$1 Incel-related} & \textbf{In next hop, 1 Incel-related} \\
\midrule
1 & $43.4\%$  & $4.1\%$   & $36.5\%$ & $2.1\%$ \\
2 & $46.5\%$  & $9.4\%$   & $38.9\%$ & $5.4\%$ \\
3 & $49.3\%$  & $11.4\%$  & $41.6\%$ & $5.0\%$ \\
4 & $49.7\%$  & $18.9\%$  & $42.0\%$ & $11.2\%$ \\
5 & $47.9\%$  & $30.1\%$  & $39.7\%$ & $17.7\%$ \\
\bottomrule
\end{tabular}%
\caption{Probability of finding (a) at least one Incel-related video in the next $5-M$ hops having already watched M consecutive Incel-related videos; and (b) an Incel-related video at hop $M+1$ assuming the user already watched $M$ consecutive Incel-related videos for both the Incel-derived and Control recommendation graphs. 
Note that in this scenario the random walker chooses to watch Incel-related videos.}
\label{tab:rabbit_hole_consecutive_hops_incel_related}
\end{table*}

\subsection{Does YouTube's recommendation algorithm contribute to steering users towards Incel communities?}
\label{subsec:random_walks_recommendation_graph}
We then study how YouTube's recommendation algorithm behaves with respect to discovering Incel-related videos.
Through our graph analysis, we showed that the problem of Incel-related videos on YouTube is quite prevalent.
However, it is still unclear how often YouTube's recommendation algorithm leads users to this type of content.

To measure this, we perform experiments considering a ``random walker.''
This allows us to simulate a random user who starts from one video and then watches several videos according to the recommendations.
More precisely, since we build our recommendation graphs using the YouTube Data API, the random walker simulates non-logged-in users who watch videos on YouTube.
The random walker begins from a randomly selected node and navigates the graph choosing edges at random for five hops. 
We repeat this process for $1,000$ random walks considering two starting scenarios. In the first scenario, the starting node is restricted to Incel-related videos. In the second, it is restricted to Other.
We perform the same experiment on both the Incel-derived and Control recommendations graphs.

Next, for the random walks of each recommendation graph, we calculate two metrics:
1) the percentage of random walks where the random walker finds at least one Incel-related video in the $k$-th hop; 
and 2) the percentage of Incel-related videos over all unique videos that the random walker encounters up to the $k$-th hop for both starting scenarios.
The two metrics, at each hop are shown in Figure~\ref{fig:random_walks_plots_at_least_one} and~\ref{fig:random_walks_plots} for both recommendation graphs.

When starting from an Other video, there is, respectively, a $10.8\%$ and $6.3\%$ probability to encounter at least one Incel-related video after five hops in the Incel-derived and Control recommendation graphs (see Figure~\ref{fig:at_least_one_random_walks_plot_started_from_irrelevant}). When starting from an Incel-related video, we find at least one Incel-related in $43.4\%$ and $36.5\%$ of the random walks performed on the Incel-derived and Control recommendation graphs, respectively (see Figure~\ref{fig:at_least_one_random_walks_plot_started_from_incels_related}). Also, when starting from Other videos, most of the Incel-related videos are found early in our random walks (i.e., at the first hop), and this number remains almost the same as the number of hops increases (see Figure~\ref{fig:random_walks_plot_started_from_irrelevant}). 
The same stands when starting from Incel-related videos, but in this case, the percentage of Incel-related videos decreases as the number of hops increases for both recommendation graphs (see Figure~\ref{fig:random_walks_plot_started_from_incels_related}).

As expected, in all cases, the probability of encountering Incel-related videos in random walks performed on the Incel-derived recommendation graph is higher than in the random walks performed on the Control recommendation graph. 
We also verify that the difference between the distribution of Incel-related videos encountered in the random walks of the two recommendation graphs is statistically significant via the Kolmogorov-Smirnov test~\cite{massey1951kolmogorov} ($p<0.05$).
Overall, we find that Incel-related videos are usually recommended within the two first hops. 
However, in subsequent hops, the number of encountered Incel-related videos decreases. 
This indicates that in the absence of personalization (e.g., for a non-logged-in user), a user casually browsing YouTube videos is unlikely to end up in a region dominated by Incel-related videos.

\subsection{Does the frequency with which Incel-related videos are recommended increase for users who choose to see the content?} 
So far, we have simulated the scenario where a user browses the recommendation graph randomly, i.e., they do {\em not} select Incel-related videos according to their interests or other cues nudging them to view certain content.
Next, we simulate the behavior of a user who chooses to watch a few Incel-related videos and investigate whether or not he will get recommended Incel-related videos with a higher probability within the next few hops.

Table~\ref{tab:rabbit_hole_consecutive_hops_incel_related} reports how likely it is for a user to encounter Incel-related videos assuming he has already watched a few. 
To do so, we use the random walks performed on the Incel-derived and Control recommendation graphs in section~\ref{subsec:random_walks_recommendation_graph}. 
We consider only the random walks started from an Incel-related video, and we zero in on those where the user watches consecutive Incel-related videos.
Specifically, we report two metrics: 1)~the probability that a user encounters at least one Incel-related video in $5-M$ hops, having already seen M consecutive Incel-related videos; 
and 2)~the probability that the user will encounter an Incel-related video on the $M+1$ hop, assuming he has already seen $M$ consecutive Incel-related videos. 
Note that, at each hop $M$ of a random walk, we calculate both metrics by only considering the random walks for which all the videos encountered in the first M hops of the walk were Incel-related.
These metrics allow us to understand whether the recommendation algorithm keeps recommending Incel-related videos to a user who starts watching a few of them.

\begin{figure}
\centering
\includegraphics[width=.7\columnwidth]{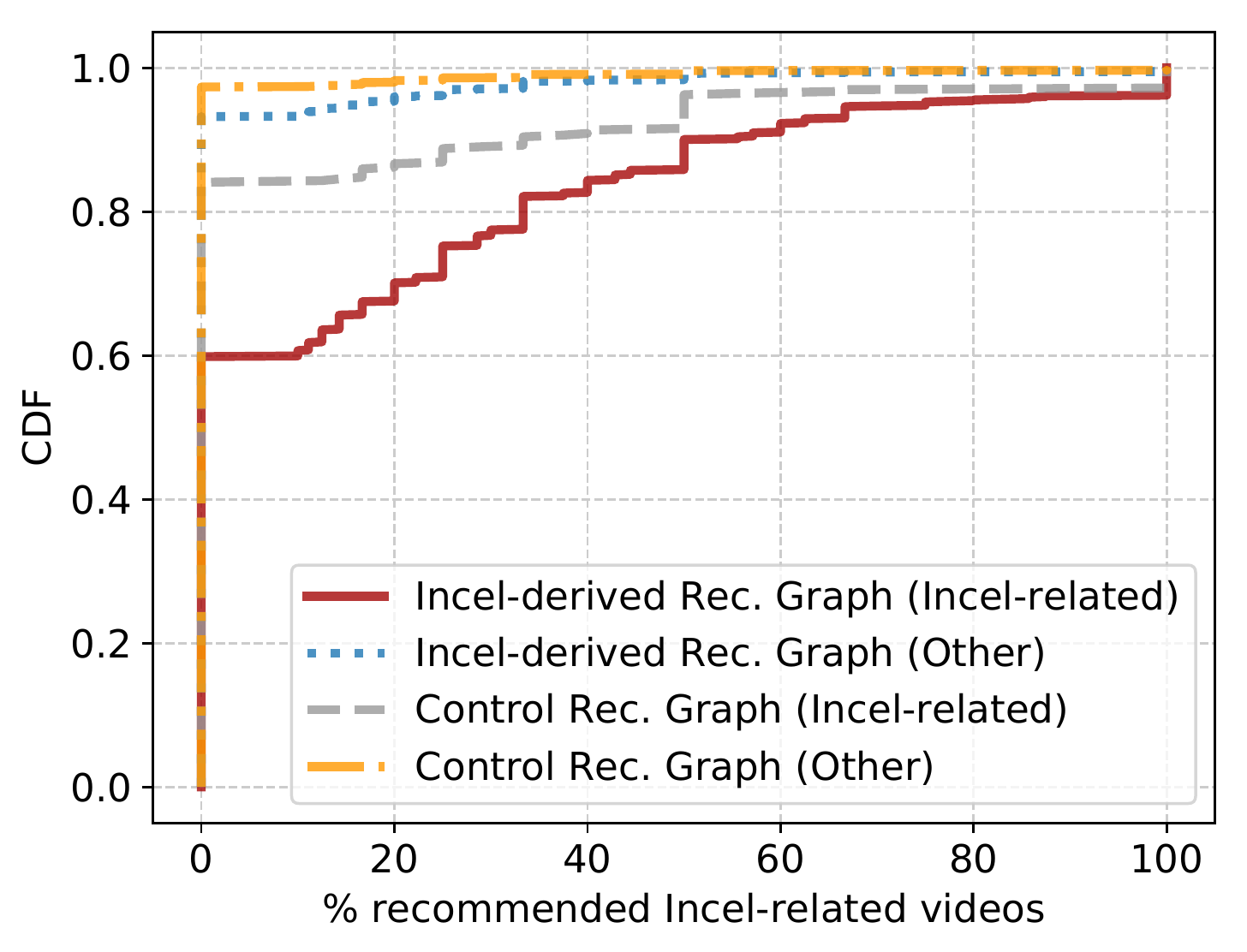}
\caption{CDF of the percentage of recommended Incel-related videos per video for both Incel-related and other videos in the Incels-derived and Control recommendation graphs.}
\label{fig:graphs_perc_recommended_incel_related_per_video_cdf}
\end{figure}

At every hop M, there is a $\geq43.4\%$ and $\geq36.5\%$ chance to encounter at least one Incel-related video within $5-M$ hops in the Incel-derived and Control recommendation graphs, respectively (second and fourth column in Table~\ref{tab:rabbit_hole_consecutive_hops_incel_related}). 
Furthermore, by looking at the probability of encountering an Incel-related video at hop $M+1$, having already watched $M$ Incel-related videos (third and right-most column in Table~\ref{tab:rabbit_hole_consecutive_hops_incel_related}), we find an increasingly higher chance as the number of consecutive Incel-related increases. 
Specifically, for the Incel-derived recommendation graph, the probability rises from $4.1\%$ at the first hop to $30.1\%$ for the last hop. For the Control recommendation graph, it rises from $2.1\%$ to $17.7\%$.

These findings unveil that as users watch Incel-related videos, the algorithm recommends other Incel-related content with increasing frequency.
In Figure~\ref{fig:graphs_perc_recommended_incel_related_per_video_cdf}, we plot the CDF of the percentage of Incel-related recommendations for each node in both recommendation graphs. 
In the Incel-derived recommendation graph, $4.6\%$ of the Incel-related videos have more than $80.0\%$ Incel-related recommendations, while $10.0\%$ of the Incel-related videos have more than $50.0\%$ Incel-related recommendations. 
The percentage of Other videos that have more than $50.0\%$ Incel-related recommendations is negligible. 
Although the percentage of Incel-related recommendations is lower, we see similar trends for the Control recommendation graph: $8.6\%$ of the Incel-related videos have more than $50.0\%$ Incel-related recommendations.

Arguably, the effect we observe may be a contributor to the anecdotally reported echo chamber effect. 
This effect entails a viewer who begins to engage with this type of content and likely falls into an algorithmic rabbit hole, with recommendations becoming increasingly dominated by such harmful content and beliefs, which also becomes increasingly extreme~\cite{cinelli2021echo,mozila2019RabbitHoleStories,youtuberadical_2019,rabbithole2019donovan,ribeiro2019auditing}. 
However, the degree to which the above-inferred algorithm characteristics contribute to a possible echo chamber effect depends on: 1) personalization factors; and 2) the ability to measure whether recommendations become increasingly extreme.

\subsection{Take-Aways}
Overall, our analysis of YouTube's recommendation algorithm yields the following main findings:
\begin{enumerate}
\item We find a non-negligible amount of Incel-related videos ($2.9\%$) within YouTube's recommendation graph being recommended to users (see Table~\ref{tab:videos_in_recommendations_rounds});
\item When a user watches a non-Incel-related video, if they randomly follow one of the top ten recommended videos, there is a $2.8\%$ chance they will end up with an Incel-related video (see Table~\ref{tab:graph_transitions_in_recommendation_rounds});
\item By performing random walks on YouTube's recommendation graph, we find that when starting from a random non-Incel-related video, there is a $6.3\%$ probability to encounter at least one Incel-related video within five hops (see Figure~\ref{fig:at_least_one_random_walks_plot_started_from_irrelevant});
\item As users choose to watch Incel-related videos, the algorithm recommends other Incel-related videos with increasing frequency (see the third and the right-most column of Table~\ref{tab:rabbit_hole_consecutive_hops_incel_related}).
\end{enumerate}

\section{Discussion}
\label{sec:discussion}
Our analysis points to an increase in Incel-related activity on YouTube over the past few years. 
More importantly, our recommendation graph analysis shows that Incel-related videos are recommended with increasing frequency to users who keep watching them. This indicates that recommendation algorithms, to an extent indeed, nudge users towards extreme content.
This section discusses our results in more detail and how they align with existing research in the area. 
We also discuss the technical challenges we faced and how we addressed them and highlight limitations.

\subsection{Challenges}
Our data collection and annotation efforts faced many challenges. 
First, there was no available dataset of YouTube videos related to the Incel community or any other Manosphere groups. 
Guided by other studies using Reddit as a source for collecting and analyzing YouTube videos~\cite{papadamou2020disturbed}, and based on evidence suggesting that Incels are particularly active on Reddit~\cite{farrell2019exploring,ribeiro2020pick}, we build our dataset by collecting videos shared on Incel-related communities on Reddit.
Second, devising a methodology for the annotation of the collected videos is not trivial. 
Due to the nature of the problem, we hypothesize that using a classifier on the video footage will not capture the various aspects of Incel-related activity on YouTube. This is because the misogynistic views of Incels may force them to heavily comment on a seemingly benign 
video (e.g., a video featuring a group of women discussing gender issues)~\cite{doring2019male}. 
Hence, we devise a methodology to detect Incel-related videos based on a lexicon of Incel-related terms that considers both the video's transcript and its comments. 

We believe that the scientific community can use our text-based approach to study other misogynistic ideologies on the platform, which tend to have their particular glossary.

\subsection{Limitations}
Our video annotation methodology might flag some benign videos as Incel-related. 
This can be a false positive or due to Incels that heavily comment on (or even raid~\cite{enrico2019cscw}) a benign video (e.g., a video featuring a group of women discussing gender issues). However, by considering the video's transcript in our video annotation methodology, we can achieve an acceptable detection accuracy that uncovers a substantial proportion of Incel-related videos (see Section~\ref{subsec:video_annotation}). 
Despite this limitation, we believe that our video annotation methodology allows us to capture and analyze various aspects of Incel-related activity on the platform. 
Another limitation of this approach is that we may miss some Incel-related videos. 
One reason for this is that the members of web-based misogynistic communities often shift or obscure their language to avoid being detected.
Notwithstanding such limitation, our approach approaches the lower bound of the Incel-related videos available in our dataset, allowing us to conclude that the implications of YouTube's recommendation algorithm on disseminating misogynistic content are at least as profound as we observe. 

Moreover, our work does not consider per-user personalization; the video recommendations we collect represent only some of the recommendation system's facets. 
More precisely, we analyze YouTube recommendations generated based on content relevance and the user base's engagement in aggregate.
However, we believe that the recommendation graphs we obtain do allow us to understand how YouTube's recommendation system is behaving {\em in our scenario.}
Also, note that a similar methodology for auditing YouTube's recommendation algorithm has been used in previous work~\cite{ribeiro2019auditing}.

\subsection{The footprint of the Incel community on YouTube}
As mentioned earlier, prior work suggests that Reddit's decision to ban subreddits did not solve the problem~\cite{chandrasekharan2017you}, as users migrated to other platforms~\cite{newell2016user,manoel_bans}. 
At the same time, other studies show that Incels are particularly active on Reddit~\cite{farrell1996myth,ribeiro2020pick}, pinpointing the need to develop methodologies that identify and characterize Manosphere-related activities on YouTube and other social media platforms. 
Realizing the threat, Reddit took measures to tackle the problem by banning several subreddits associated with the Incel community and the Manosphere in general. 
Driven by that, we set out to study the evolution of the Incel community, over the last decade, on other platforms like YouTube.

Our results show that Incel-related activity on YouTube increased over the past few years, in particular, concerning the publication of Incel-related videos, as well as in comments that include pertinent terms. This indicates that Incels are increasingly exploiting YouTube to spread their ideology and express their misogynistic views. 
Although we do not know whether these users are banned Reddit users that migrated to YouTube or whether this increase in Incel-related activity is associated with the increased interest in Incel-related communities on Reddit over the past few years, our findings are still worrisome.
Also, Reddit's decision to ban /r/Incels for inciting violence against women~\cite{incelssubredditbanned_2017} and the observed sharp increase in Incel-related activity on YouTube after this period aligns with the theoretical framework proposed by Chandrasekharan et al.~\cite{chandrasekharan2017you}.
The increase in Incel-related activity also indicates that Reddit's decision may have energized the community and made its members more persistent. 

Despite YouTube's attempts to tackle hate~\cite{tacklehate2019youtube}, our results show that the threat is clear and present.
Also, considering that the Incel ideology is often associated with misogyny, and anti-feminist views, as well as with multiple mass murders and violent offenses~\cite{malesupr2019splc,fifthestate2019cbc}, we urge that YouTube develops effective content moderation strategies to tackle misogynistic content on the platform.

\subsection{The role of YouTube's recommendation algorithm in steering users towards the Incel community}
Driven by the fact that members of the Incel community are prone to radicalization~\cite{cecco_2019} and that YouTube has been repeatedly accused of contributing to user radicalization and promoting offensive content~\cite{rabbithole2019donovan,kaiser2018unite}, we set out to assess whether YouTube's recommendation algorithm nudges users towards Incel communities.
Using graph analysis, we analyze snapshots of YouTube's recommendation graph, finding that there is a non-negligible amount of Incel-related content being suggested to users. 
Also, by simulating a user who casually browses YouTube, we see a high chance that a user will encounter at least one Incel-related video five hops after he starts from a non-Incel-related video. 
Next, we simulate a user who, upon encountering an Incel-related video, becomes interested in this content and purposefully starts watching these types of videos. 
We do this to determine whether YouTube's recommendation graph steers such users into regions where a substantial portion of the recommended videos are Incel-related.
Once users enter such regions, they are \emph{likely} to consider such content as increasingly legitimate as they experience social proof of these narratives. They may find it difficult to escape to more benign content~\cite{youtuberadical_2019}. 
Interestingly, we find that once a user follows Incel-related videos, the algorithm recommends other Incel-related videos to him with increasing frequency. 
Our results point to the echo chamber effect~\cite{media2021echochamber,cinelli2021echo}. 
However, the echo chamber effect definition includes the notion that the extremist nature of the improper videos increases along with the frequency with which they are recommended. 
Since we do not assess whether the videos suggested in subsequent hops are becoming increasingly extreme, we cannot conclude that we find a statistically significant indication of this effect. 
Nevertheless, even if we do not find strong evidence of an echo chamber, our findings are worrisome especially when considering the extreme misogynistic beliefs of the Incel community.

To mitigate the harm caused to users by certain recommended videos and to incorporate community well-being into the objectives of its recommendation algorithm~\cite{stray2020aligning}, YouTube introduced ``user satisfaction'' metrics as input to the recommendation algorithm~\cite{zhao2019recommending}.
However, our findings show that misogynistic and harmful content is still being recommended to users and the recommendation algorithm is not able to discern and marginalize such content.
Hence, we believe that more effort is required by researchers and platforms to effectively detect and suppress such content in a proactive and timely manner.

\subsection{Design Implications}
Prior work has shown apparent user migration to increasingly extreme subcommunities within the Manosphere on Reddit~\cite{ribeiro2020pick}, and indications that YouTube recommendations serve as a pathway to radicalization. When taken along with our results, a more complete picture with respect to online extremist communities begins to emerge.

Radicalization and online extremism is clearly a \emph{multi-platform} problem. Social media platforms like Reddit, designed to allow organic creation and discovery of subcommunities, play a role, and so do platforms with algorithmic content recommendation systems.
The immediate take away is that while the radicalization process and the spread of extremist content generalize (at least to some extent) across different online extremist communities, the specific mechanism likely does not generalize across different platforms, which has implications for the design of moderation systems and strategies.

In particular, it implies that platform oriented-solutions should not exist in a vacuum, and indeed it is quite likely that information sharing between platforms could bolster overall effectiveness.
For example, an approach that could benefit both platforms we study involves using Reddit activity to help tune the YouTube recommendation algorithm and using information from the recommendation algorithm to help Reddit perform content moderation. 
In such a hypothetical arrangement, Reddit, whose content moderation team is intimately familiar with the troublesome communities, could help YouTube understand how the content these communities consume fits within the recommendation graph. 
Similarly, Reddit's moderation efforts could be bolstered with information from the YouTube recommendation graph. 
The discovery of emerging dangerous communities could be aided by understanding where the content posted by them fits within the YouTube recommendation graph compared to the content posted by known troublesome communities.

At the same time, our findings suggest that researchers who study radicalization and online extremism can benefit by performing cross-platform analysis as studying across multiple platforms can help in better understanding the footprint and evolvement of emerging dangerous communities.

\subsection{Future Work} 
We plan to extend our work by studying other Manosphere communities on YouTube (e.g., Men Going Their Own Way) and user migration between Manosphere and other reactionary communities.
We also plan to implement crawlers that will allow us to simulate real users and perform random walks on YouTube with user personalization.
This will enable measurements of YouTube’s recommendation graph while also assessing the effect of various personalization factors (e.g., gender, a user's watch history, etc.) on the amount of misogynistic content being recommended to a user.
Note that this task is not straightforward as it requires understanding and replicating multiple meaningful characteristics of Incels' behavior. 

Another interesting direction for future research is to perform a survey study on YouTube with real users and even collecting their qualitative feedback.
Last, an important direction for future work is to study the effect of the COVID-19 pandemic on the growth of web-based misogynistic communities.

\section{Conclusion}

This paper presented a large-scale data-driven characterization of the Incel community on YouTube.
We collected 6.5K YouTube videos shared by users in Incel-related communities within Reddit. We used them to understand how Incel ideology spreads on YouTube and study the evolution of the community.  
We found a non-negligible growth in Incel-related activity on YouTube over the past few years, both in terms of Incel-related videos published and comments likely posted by Incels. This result suggests that users gravitating around the Incel community are increasingly using YouTube to disseminate their views.

Overall, our study is a first step towards understanding the Incel community and other misogynistic ideologies on YouTube. 
We argue that it is crucial to protect potential radicalization ``victims'' by developing methods and tools to detect Incel-related videos and other misogynistic activities on YouTube.  Our analysis shows growth in Incel-related activities on Reddit and highlights how the Incel community operates on multiple platforms and Web communities. This also prompts the need to perform more multi-platform studies to understand Manosphere communities further.

We also analyzed how YouTube's recommendation algorithm behaves with respect to Incel-related videos. 
By performing random walks on the recommendation graph, we estimated a $6.3\%$ chance for a user who starts by watching non-Incel-related videos to be recommended Incel-related ones within five recommendation hops.
At the same time, users who have seen two or three Incel-related videos at the start of their walk see recommendations that consist of $9.4\%$ and $11.4\%$ Incel-related videos, respectively. Moreover, the portion of Incel-related recommendations increases substantially as the user watches an increasing number of consecutive Incel-related videos. 

Our results highlight the pressing need to further study and understand the role of YouTube's recommendation algorithm in users' radicalization and content consumption patterns.  Ideally, a recommendation algorithm should avoid recommending potentially harmful or extreme videos. However, our analysis confirms prior work showing that this is not always the case on YouTube~\cite{ribeiro2019auditing}.

\descr{Acknowledgments.}
This project has received funding from the European Union's Horizon 2020 Research and Innovation program under the Marie Sk\l{}dowska-Curie ENCASE project (GA No. 691025) and the CONCORDIA project (GA No. 830927), the US National Science Foundation (grants: 1942610, 2114407, 2114411, and 2046590), and the UK's National Research Centre on Privacy, Harm Reduction, and Adversarial Influence Online (UKRI grant: EP/V011189/1).
This work reflects only the authors' views; the funding agencies are not responsible for any use that may be made of the information it contains.

\small
\bibliographystyle{abbrv}

\begin{thebibliography}{10}

\bibitem{gdpr2018rightaccess}
{GDPR: Right of Access}.
\newblock \url{https://gdpr-info.eu/issues/right-of-access/}, 2018.

\bibitem{gdpr2018rightforgotten}
{GDPR: Right to be Forgotten}.
\newblock \url{https://gdpr-info.eu/issues/right-to-be-forgotten/}, 2018.

\bibitem{agarwal2014focused}
S.~Agarwal and A.~Sureka.
\newblock {A Focused Crawler for Mining Hate and Extremism Promoting Videos on
  YouTube}.
\newblock In {\em Proceedings of the 25th ACM conference on Hypertext and
  social media}, 2014.

\bibitem{baele2019saint}
S.~J. Baele, L.~Brace, and T.~G. Coan.
\newblock {From "Incel" to "Saint": Analyzing the violent worldview behind the
  2018 Toronto attack}.
\newblock In {\em Terrorism and Political Violence}. Routledge, 2019.

\bibitem{bashir2018doing}
N.~Bashir.
\newblock {Doing Research in Peoples’ Homes: Fieldwork, Ethics and Safety--On
  the Practical Challenges of Researching and Representing Life on the
  Margins}.
\newblock In {\em Qualitative Research}, 2018.

\bibitem{baumgartner2020pushshift}
J.~Baumgartner, S.~Zannettou, B.~Keegan, M.~Squire, and J.~Blackburn.
\newblock {The Pushshift Reddit Dataset}.
\newblock In {\em Proceedings of the International AAAI Conference on Web and
  Social Media}, 2020.

\bibitem{HowRampageKiller2018}
{BBC}.
\newblock How rampage killer became misogynist ``hero''.
\newblock \url{https://www.bbc.com/news/world-us-canada-43892189}, 2018.

\bibitem{beale2004impact}
B.~Beale, R.~Cole, S.~Hillege, R.~McMaster, and S.~Nagy.
\newblock {Impact of In-depth Interviews on the Interviewer: Roller Coaster
  Ride}.
\newblock In {\em Nursing \& health sciences}, 2004.

\bibitem{incel2018vox}
Z.~Beauchamp.
\newblock {Incel, the misogynist ideology that inspired the deadly Toronto
  attack, explained}.
\newblock
  \url{https://www.vox.com/world/2018/4/25/17277496/incel-toronto-attack-alek-minassian},
  2018.

\bibitem{blagden2010challenge}
N.~Blagden and S.~Pemberton.
\newblock {The Challenge in Conducting Qualitative Research With Convicted Sex
  Offenders}.
\newblock In {\em The Howard Journal of Criminal Justice}, 2010.

\bibitem{blais2012masculinism}
M.~Blais and F.~Dupuis-D{\'e}ri.
\newblock {Masculinism and the Antifeminist Countermovement}.
\newblock In {\em Social Movement Studies}. Taylor \& Francis, 2012.

\bibitem{bratich2019pickup}
J.~Bratich and S.~Banet-Weiser.
\newblock {From Pick-Up Artists to Incels: Con(fidence) Games, Networked
  Misogyny, and the Failure of Neoliberalism}.
\newblock In {\em International Journal of Communication}, 2019.

\bibitem{cecco_2019}
L.~Cecco.
\newblock {Toronto van attack suspect says he was 'radicalized' online by
  'incels'}.
\newblock
  \url{https://www.theguardian.com/world/2019/sep/27/alek-minassian-toronto-van-attack-interview-incels},
  2019.

\bibitem{malesupr2019splc}
S.~P.~L. Center.
\newblock {Male Supremacy}.
\newblock
  \url{https://www.splcenter.org/fighting-hate/extremist-files/ideology/male-supremacy},
  2019.

\bibitem{chandrasekharan2017you}
E.~Chandrasekharan, U.~Pavalanathan, A.~Srinivasan, A.~Glynn, J.~Eisenstein,
  and E.~Gilbert.
\newblock {You Can't Stay Here: The Efficacy of Reddit's 2015 Ban Examined
  Through Hate Speech}.
\newblock In {\em Proceedings of the ACM on Human-Computer Interaction}, number
  CSCW. ACM New York, NY, USA, 2017.

\bibitem{cinelli2021echo}
M.~Cinelli, G.~D.~F. Morales, A.~Galeazzi, W.~Quattrociocchi, and M.~Starnini.
\newblock {The Echo Chamber Effect on Social Media}.
\newblock In {\em Proceedings of the National Academy of Sciences}, 2021.

\bibitem{vice2018incellangdecode}
A.~Conti.
\newblock {Learn to Decode the Secret Language of the Incel Subculture}.
\newblock
  \url{https://www.vice.com/en/article/7xmaze/learn-to-decode-the-secret-language-of-the-incel-subculture},
  2018.

\bibitem{incelsLooksmaxing2018}
J.~Cook.
\newblock {Inside Incels' Looksmaxing Obsession: Penis Stretching, Skull
  Implants And Rage}.
\newblock
  \url{https://www.huffpost.com/entry/incels-looksmaxing-obsession_n_5b50e56ee4b0de86f48b0a4f},
  2018.

\bibitem{covington2016deep}
P.~Covington, J.~Adams, and E.~Sargin.
\newblock {Deep Neural Networks for YouTube Recommendations}.
\newblock In {\em Proceedings of the 10th ACM conference on recommender
  systems}, 2016.

\bibitem{dittrich2012menlo}
D.~Dittrich, E.~Kenneally, et~al.
\newblock {The Menlo Report: Ethical Principles Guiding Information and
  Communication Technology Research}.
\newblock Technical report, US Department of Homeland Security, 2012.

\bibitem{doring2019male}
N.~D{\"o}ring and M.~R. Mohseni.
\newblock {Male Dominance and Sexism on YouTube: Results of Three Content
  Analyses}.
\newblock In {\em Feminist Media Studies}. Taylor \& Francis, 2019.

\bibitem{fifthestate2019cbc}
T.~F. Estate.
\newblock {Why incels are a 'real and present threat' for Canadians}.
\newblock
  \url{https://www.cbc.ca/news/canada/incel-threat-canadians-fifth-estate-1.4992184},
  2019.

\bibitem{farrell2019exploring}
T.~Farrell, M.~Fernandez, J.~Novotny, and H.~Alani.
\newblock {Exploring Misogyny Across the Manosphere in Reddit}.
\newblock In {\em Proceedings of the 10th ACM Conference on Web Science}, 2019.

\bibitem{farrell1996myth}
W.~Farrell.
\newblock {\em {The Myth of Male Power}}.
\newblock Berkeley Publishing Group, 1996.

\bibitem{fisher1922interpretation}
R.~A. Fisher.
\newblock {On the Interpretation of $\chi$ 2 From Contingency Tables, and the
  Calculation of P}.
\newblock {\em Journal of the Royal Statistical Society}, 1922.

\bibitem{fleiss1971measuring}
J.~L. Fleiss.
\newblock {Measuring Nominal Scale Agreement Among Many Raters}.
\newblock In {\em Psychological bulletin}, 1971.

\bibitem{giannakopoulos2010multimodal}
T.~Giannakopoulos, A.~Pikrakis, and S.~Theodoridis.
\newblock {A Multimodal Approach to Violence Detection in Video Sharing Sites}.
\newblock In {\em 2010 20th International Conference on Pattern Recognition}.
  IEEE, 2010.

\bibitem{ging2019alphas}
D.~Ging.
\newblock {Alphas, Betas, and Incels: Theorizing the Masculinities of the
  Manosphere}.
\newblock In {\em Men and Masculinities}. SAGE Publications Sage CA: Los
  Angeles, CA, 2019.

\bibitem{youtubedataapi_2019}
{Google Developers}.
\newblock {YouTube Data API}.
\newblock \url{https://developers.google.com/youtube/v3/}, 2020.

\bibitem{gotell2016SexualViolenceManospherea}
L.~Gotell and E.~Dutton.
\newblock {Sexual Violence in the ``Manosphere'': Antifeminist Men's Rights
  Discourses on Rape}.
\newblock In {\em International Journal for Crime, Justice and Social
  Democracy}. Queensland University of Technology, 2016.

\bibitem{incelssubredditbanned_2017}
C.~Hauser.
\newblock Reddit bans "incel" group for inciting violence against women.
\newblock
  \url{https://www.nytimes.com/2017/11/09/technology/incels-reddit-banned.html},
  2017.

\bibitem{hoffman2020assessing}
B.~Hoffman, J.~Ware, and E.~Shapiro.
\newblock {Assessing the Threat of Incel Violence}.
\newblock In {\em Studies in Conflict \& Terrorism}. Taylor \& Francis, 2020.

\bibitem{hunteFemaleNatureCucks}
Z.~Hunte and K.~Engstr{\"o}m.
\newblock {``Female Nature, Cucks, and Simps'': Understanding Men Going Their
  Own Way as Part of the Manosphere}.
\newblock Master's thesis, 2019.

\bibitem{hussein2020measuring}
E.~Hussein, P.~Juneja, and T.~Mitra.
\newblock {Measuring Misinformation in Video Search Platforms: An Audit Study
  on YouTube}.
\newblock {\em Proceedings of the ACM on Human-Computer Interaction}, (CSCW),
  2020.

\bibitem{incelwikiBlackpillIncelWiki}
{Incels Wiki}.
\newblock Blackpill.
\newblock \url{https://incels.wiki/w/Blackpill}, 2019.

\bibitem{incelglossary_2019}
{Incels Wiki}.
\newblock {Incel Forums Term Glossary}.
\newblock \url{https://incels.wiki/w/Incel_Forums_Term_Glossary}, 2019.

\bibitem{incelswiki_2019}
{Incels Wiki}.
\newblock {The Incel Wiki}.
\newblock \url{https://incels.wiki}, 2019.

\bibitem{jaki2019online}
S.~Jaki, T.~De~Smedt, M.~Gw{\'o}{\'z}d{\'z}, R.~Panchal, A.~Rossa, and
  G.~De~Pauw.
\newblock {Online Hatred of Women in the Incels.me Forum: Linguistic Analysis
  and Automatic Detection}.
\newblock In {\em Journal of Language Aggression and Conflict}, 2019.

\bibitem{jiang2019bias}
S.~Jiang, R.~E. Robertson, and C.~Wilson.
\newblock {Bias Misperceived: The Role of Partisanship and Misinformation in
  YouTube Comment Moderation}.
\newblock In {\em Proceedings of the International AAAI Conference on Web and
  Social Media}, 2019.

\bibitem{kaiser2018unite}
J.~Kaiser and A.~Rauchfleisch.
\newblock Unite the right? how youtube’s recommendation algorithm connects
  the us far-right.
\newblock In {\em D\&S Media Manipulation}, 2018.

\bibitem{kiernan1988remains}
K.~E. Kiernan.
\newblock {Who Remains Celibate?}
\newblock In {\em Journal of Biosocial Science}. Cambridge University Press,
  1988.

\bibitem{landis1977MeasurementObserverAgreementa}
J.~R. Landis and G.~G. Koch.
\newblock {The Measurement of Observer Agreement for Categorical Data}.
\newblock In {\em Biometrics}. JSTOR, 1977.

\bibitem{linAntifeminismOnlineMGTOW2017}
J.~L. Lin.
\newblock {Antifeminism {Online}: {MGTOW} ({Men} {Going} {Their} {Own} {Way})}.
\newblock JSTOR, 2017.

\bibitem{media2021echochamber}
D.~M. Literacy.
\newblock {What is an echo chamber?}
\newblock
  \url{https://edu.gcfglobal.org/en/digital-media-literacy/what-is-an-echo-chamber/1/},
  2021.

\bibitem{enrico2019cscw}
E.~Mariconti, G.~Suarez-Tangil, J.~Blackburn, E.~De~Cristofaro, N.~Kourtellis,
  I.~Leontiadis, J.~L. Serrano, and G.~Stringhini.
\newblock {``You Know What to Do'': Proactive Detection of YouTube Videos
  Targeted by Coordinated Hate Attacks}.
\newblock In {\em Proceedings of the ACM on Human-Computer Interaction}, number
  CSCW. ACM New York, NY, USA, 2019.

\bibitem{youtubeBansExtremism2019}
P.~Martineau.
\newblock {YouTube Is Banning Extremist Videos. Will It Work?}
\newblock
  \url{https://www.wired.com/story/how-effective-youtube-latest-ban-extremism/},
  2019.

\bibitem{marwick2017media}
A.~Marwick and R.~Lewis.
\newblock {Media Manipulation and Disinformation Online}.
\newblock 2017.

\bibitem{massanari2017gamergate}
A.~Massanari.
\newblock {\# Gamergate and The Fappening: How Reddit’s Algorithm,
  Governance, and Culture Support Toxic Technocultures}.
\newblock In {\em New Media \& Society}. Sage Publications Sage UK: London,
  England, 2017.

\bibitem{massey1951kolmogorov}
F.~J. Massey~Jr.
\newblock {The Kolmogorov-Smirnov Test for Goodness of Fit}.
\newblock In {\em Journal of the American statistical Association}. Taylor \&
  Francis, 1951.

\bibitem{maxwell2020short}
D.~Maxwell, S.~R. Robinson, J.~R. Williams, and C.~Keaton.
\newblock {"A Short Story of a Lonely Guy": A Qualitative Thematic Analysis of
  Involuntary Celibacy Using Reddit}.
\newblock In {\em Sexuality \& Culture}. Springer, 2020.

\bibitem{menzie2020stacys}
L.~Menzie.
\newblock {Stacys, Beckys, and Chads: The Construction of Femininity and
  Hegemonic Masculinity Within Incel Rhetoric}.
\newblock In {\em Psychology \& Sexuality}. Taylor \& Francis, 2020.

\bibitem{messner1998limits}
M.~A. Messner.
\newblock {The Limits of “The Male Sex Role” An Analysis of the Men's
  Liberation and Men's Rights Movements' Discourse}.
\newblock In {\em Gender \& Society}. SAGE Publications, Inc., 1998.

\bibitem{mozila2019RabbitHoleStories}
{Mozila Foundation}.
\newblock Got youtube regrets? thousands do!
\newblock \url{https://foundation.mozilla.org/en/campaigns/youtube-regrets/},
  2019.

\bibitem{nagle2015investigation}
A.~Nagle.
\newblock {\em {An Investigation into Contemporary Online Anti-feminist
  Movements}}.
\newblock PhD thesis, Dublin City University, 2015.

\bibitem{newell2016user}
E.~Newell, D.~Jurgens, H.~Saleem, H.~Vala, J.~Sassine, C.~Armstrong, and
  D.~Ruths.
\newblock {User Migration in Online Social Networks: A case study on Reddit
  during a period of Community Unrest}.
\newblock In {\em Proceedings of the International AAAI Conference on Web and
  Social Media}, 2016.

\bibitem{rabbithole2019donovan}
C.~O'Donovan, C.~Warzel, L.~McDonald, B.~Clifton, and M.~Woolf.
\newblock {We Followed YouTube's Recommendation Algorithm Down The Rabbit
  Hole}.
\newblock
  \url{https://www.buzzfeednews.com/article/carolineodonovan/down-youtubes-recommendation-rabbithole},
  2019.

\bibitem{aoir2019ethicalguidelines}
A.~of~Internet~Research.
\newblock {Internet Research: Ethical Guidelines 3.0}.
\newblock \url{https://aoir.org/reports/ethics3.pdf}, 2019.

\bibitem{insideincels2018washington}
A.~Ohlheiser.
\newblock {Inside the online world of "incels," the dark corner of the Internet
  linked to the Toronto suspect}.
\newblock \url{https://tinyurl.com/incels-link-toronto-suspect}, 2018.

\bibitem{ottoni2018analyzing}
R.~Ottoni, E.~Cunha, G.~Magno, P.~Bernardina, W.~Meira~Jr, and V.~Almeida.
\newblock {Analyzing Right-wing YouTube Channels: Hate, Violence and
  Discrimination}.
\newblock In {\em Proceedings of the 10th ACM Conference on Web Science}, 2018.

\bibitem{papadamou2020disturbed}
K.~Papadamou, A.~Papasavva, S.~Zannettou, J.~Blackburn, N.~Kourtellis,
  I.~Leontiadis, G.~Stringhini, and M.~Sirivianos.
\newblock {Disturbed YouTube for Kids: Characterizing and Detecting
  Inappropriate Videos Targeting Young Children}.
\newblock In {\em Proceedings of the International AAAI Conference on Web and
  Social Media}, 2020.

\bibitem{pariser2011filter}
E.~Pariser.
\newblock {\em {The Filter Bubble: How the New Personalized Web is Changing
  What We Read and How We Think}}.
\newblock Penguin, 2011.

\bibitem{rationalwikiIncelRationalWiki}
{Rational Wiki}.
\newblock Incel.
\newblock \url{https://rationalwiki.org/wiki/Incel}, 2019.

\bibitem{ribeiro2020pick}
M.~H. Ribeiro, J.~Blackburn, B.~Bradlyn, E.~De~Cristofaro, G.~Stringhini,
  S.~Long, S.~Greenberg, and S.~Zannettou.
\newblock {The Evolution of the Manosphere Across the Web}.
\newblock In {\em Proceedings of the International AAAI Conference on Web and
  Social Media}, 2021.

\bibitem{manoel_bans}
M.~H. Ribeiro, S.~Jhaver, S.~Zannettou, J.~Blackburn, E.~De~Cristofaro,
  G.~Stringhini, and R.~West.
\newblock {Does Platform Migration Compromise Content Moderation? Evidence from
  r/The\_Donald and r/Incels}.
\newblock In {\em arXiv preprint 2010.10397}, 2020.

\bibitem{ribeiro2019auditing}
M.~H. Ribeiro, R.~Ottoni, R.~West, V.~A. Almeida, and W.~Meira~Jr.
\newblock {Auditing Radicalization Pathways on YouTube}.
\newblock In {\em Proceedings of the 2020 Conference on Fairness,
  Accountability, and Transparency}, 2020.

\bibitem{rivers2014ethical}
C.~M. Rivers and B.~L. Lewis.
\newblock {Ethical Research Standards in a World of Big Data}.
\newblock In {\em F1000Research}, volume~3, 2014.

\bibitem{theverge2019raincelsban}
A.~Robertson.
\newblock Reddit has broadened its anti-harassment rules and banned a major
  incel forum.
\newblock
  \url{https://www.theverge.com/2019/9/30/20891920/reddit-harassment-bullying-threats-new-policy-change-rules-subreddits},
  2019.

\bibitem{youtuberadical_2019}
K.~Roose.
\newblock {The Making of a YouTube Radical}.
\newblock \url{https://tinyurl.com/nytimes-youtube-radical}, 2019.

\bibitem{harpermercer_2015}
S.~Sara, K.~Lah, S.~Almasy, and R.~Ellis.
\newblock Oregon shooting: {{Gunman}} a student at {{Umpqua Community
  College}}.
\newblock \url{https://tinyurl.com/umpqua-community-college-shoot}, 2015.

\bibitem{stocker2020riding}
C.~St{\"o}cker and M.~Preuss.
\newblock {Riding the Wave of Misclassification: How We End up with Extreme
  YouTube Content}.
\newblock In {\em International Conference on Human-Computer Interaction}.
  Springer, 2020.

\bibitem{stray2020aligning}
J.~Stray.
\newblock Aligning ai optimization to community well-being.
\newblock {\em International Journal of Community Well-Being}, 3(4):443--463,
  2020.

\bibitem{sureka2010mining}
A.~Sureka, P.~Kumaraguru, A.~Goyal, and S.~Chhabra.
\newblock {Mining YouTube to Discover Extremist Videos, Users and Hidden
  Communities}.
\newblock In {\em Asia Information Retrieval Symposium}. Springer, 2010.

\bibitem{tahir2019BringingKidBacka}
R.~Tahir, F.~Ahmed, H.~Saeed, S.~Ali, F.~Zaffar, and C.~Wilson.
\newblock Bringing the {{Kid}} back into {{YouTube Kids}}: {{Detecting
  Inappropriate Content}} on {{Video Streaming Platforms}}.
\newblock In {\em IEEE/ACM International Conference on Advances in Social
  Networks Analysis and Mining (ASONAM)}. IEEE, 2019.

\bibitem{tacklehate2019youtube}
T.~Y. Team.
\newblock {Our ongoing work to tackle hate}.
\newblock
  \url{https://blog.youtube/news-and-events/our-ongoing-work-to-tackle-hate},
  2019.

\bibitem{tufekci2018youtube}
Z.~Tufekci.
\newblock {YouTube, The Great Radicalizer}.
\newblock In {\em The New York Times}, 2018.

\bibitem{vijaymeena2016survey}
M.~Vijaymeena and K.~Kavitha.
\newblock {A Survey on Similarity Measures in Text Mining}.
\newblock In {\em Machine Learning and Applications: An International Journal
  (MLAIJ)}, 2016.

\bibitem{plymouth}
M.~Weaver and S.~Morris.
\newblock {Plymouth gunman: a hate-filled misogynist and `Incel'}.
\newblock
  \url{https://www.theguardian.com/uk-news/2021/aug/13/plymouth-shooting-suspect-what-we-know-jake-davison},
  2021.

\bibitem{elliotrodgers_2014}
Wikipedia.
\newblock 2014 isla vista killings.
\newblock \url{https://en.wikipedia.org/wiki/2014_Isla_Vista_killings}, 2019.

\bibitem{wotanis2014performing}
L.~Wotanis and L.~McMillan.
\newblock {Performing Gender on YouTube: How Jenna Marbles Negotiates a Hostile
  Online Environment}.
\newblock Taylor \& Francis, 2014.

\bibitem{zannettou2018good}
S.~Zannettou, S.~Chatzis, K.~Papadamou, and M.~Sirivianos.
\newblock {The Good, the Bad and the Bait: Detecting and Characterizing
  Clickbait on YouTube}.
\newblock In {\em 2018 IEEE Security and Privacy Workshops (SPW)}. IEEE, 2018.

\bibitem{zhao2019recommending}
Z.~Zhao, L.~Hong, L.~Wei, J.~Chen, A.~Nath, S.~Andrews, A.~Kumthekar,
  M.~Sathiamoorthy, X.~Yi, and E.~Chi.
\newblock {Recommending What Video to Watch Next: A Multitask Ranking System}.
\newblock In {\em Proceedings of the 13th ACM Conference on Recommender
  Systems}, 2019.

\bibitem{zimmerman2018recognizing}
S.~Zimmerman, L.~Ryan, and D.~Duriesmith.
\newblock Recognizing the violent extremist ideology of 'incels'.
\newblock In {\em Women in International Security Policy}, 2018.

\end{thebibliography}

\end{document}